\documentclass[aps,preprint,nofootinbib,showpacs]{revtex4}
\usepackage{graphicx,color,amsmath}
\newcommand{\be}{\begin{equation}}
\newcommand{\ee}{\end{equation}}
\newcommand{\ba}{\begin{eqnarray}}
\newcommand{\beq}{\begin{equation}}
\newcommand{\eeq}{\end{equation}}
\newcommand{\ea}{\end{eqnarray}}

\newcommand{\GeV}{\text{GeV}}
\newcommand{\TeV}{\text{TeV}}

\def\beqa{\begin{eqnarray}}
\def\eeqa{\end{eqnarray}}

\def\bea{\begin{eqnarray}}
\def\eea{\end{eqnarray}}

\def\err#1#2{\lower2pt\hbox{ $\stackrel{\scriptstyle +#1}{\scriptstyle -#2}$}}
\def\ga{\mathrel{\raise.3ex\hbox{$>$\kern-.75em\lower1ex\hbox{$\sim$}}}}
\def\la{\mathrel{\raise.3ex\hbox{$<$\kern-.75em\lower1ex\hbox{$\sim$}}}}
\def\bmaT{\left(\begin{array}{ccc}}
\def\emaT{\end{array}\right)}
\def\bma{\left( \begin{array} }
\def\ema{\end{array} \right)}
\def\gsim{~{\rlap{\lower 3.5pt\hbox{$\mathchar\sim$}}\raise 1pt\hbox{$>$}}\,}
\def\lsim{~{\rlap{\lower 3.5pt\hbox{$\mathchar\sim$}}\raise 1pt\hbox{$<$}}\,}

\begin{document}

\preprint{
\vbox{%
\hbox{SHEP-11-10}
}}
\title{\boldmath Production of doubly charged scalars from the decay 
of a \\heavy SM-like Higgs boson in the Higgs Triplet Model\unboldmath} 
\author{A.G. Akeroyd}
\email{a.g.akeroyd@soton.ac.uk}
\author{S. Moretti}
\email{S.Moretti@soton.ac.uk}
\affiliation{$^{1}$School of Physics and Astronomy, University of Southampton, \\
Highfield, Southampton SO17 1BJ, United Kingdom,}
\affiliation{$^{2}$Particle Physics Department, Rutherford Appleton Laboratory, Chilton, Didcot, 
Oxon OX11 0QX, United Kingdom}

\begin{abstract}

The Higgs Triplet Model (HTM) of neutrino mass generation predicts the existence of doubly charged Higgs bosons ($H^{\pm\pm}$).
In the HTM a scalar eigenstate ($H_2$) is dominantly composed of
the scalar field from the isospin doublet, and could be significantly heavier than $H^{\pm\pm}$. 
Such a scenario would allow the possibility of a large branching ratio for the decay
$H_2\to H^{++}H^{--}$. From the production mechanism of gluon-gluon fusion, $gg\to H_2$, the above decay
mode would give rise to pair production of doubly charged Higgs bosons 
($H^{++}H^{--}$) with a cross section which could be significantly larger than the cross sections 
for the standard production mechanisms $q\overline q \to \gamma,Z\to H^{++}H^{--}$
and $q'\overline q \to W \to H^{\pm\pm}H^{\mp}$. We discuss the phenomenological
consequences for the ongoing searches for $H^{\pm\pm}$ at the Tevatron and at the LHC.

\end{abstract}
\pacs{14.80.Fd, 12.60.Fr}
\maketitle


\section{Introduction} 
The established evidence that neutrinos oscillate and possess a small mass
below the electron volt (eV) scale~\cite{Fukuda:1998mi}
necessitates physics beyond the Standard Model
(SM), which could manifest itself at the CERN Large Hadron Collider (LHC) and the Fermilab Tevatron, 
and/or in low energy experiments which 
search for lepton flavour violation (LFV)~\cite{Kuno:1999jp}.
Consequently, models of neutrino mass generation which can be probed
at present and forthcoming experiments are of great phenomenological
interest. 

Neutrinos may obtain mass via the vacuum expectation value
(vev) of a neutral Higgs boson in an isospin triplet
representation~\cite{Konetschny:1977bn, Mohapatra:1979ia, Magg:1980ut,Schechter:1980gr,Cheng:1980qt}.
A particularly simple implementation of this
mechanism of neutrino mass generation is the 
``Higgs Triplet Model'' (HTM)
in which the SM Lagrangian is augmented solely by
a $SU(2)$ triplet of scalar particles $\Delta$ 
with hypercharge $Y=2$~\cite{Konetschny:1977bn, Schechter:1980gr,Cheng:1980qt}.
In the HTM, neutrinos acquire a Majorana mass
given by the product of a triplet Yukawa coupling ($h_{ij}$,  with
$i,j=e,\mu,\tau$)
and a triplet vev ($v_\Delta$).
Consequently, there is a direct connection 
between $h_{ij}$ and the neutrino mass matrix, which gives rise to 
phenomenological predictions for processes which depend on $h_{ij}$.
A distinctive signal of the HTM would be the observation of doubly charged Higgs bosons ($H^{\pm\pm}$)
whose mass ($M_{H^{\pm\pm}}$) may be of the order of the electroweak scale.
 Such particles 
could be produced with sizeable rates at hadron colliders via the
processes $q\overline q\to H^{++}H^{--}$ \cite{Barger:1982cy, Gunion:1989in, Muhlleitner:2003me, Han:2007bk, Huitu:1996su}
and $q'\overline q \to H^{\pm\pm}H^{\mp}$~\cite{Barger:1982cy, Dion:1998pw, Akeroyd:2005gt}.
The first searches for $H^{\pm\pm}$ at a hadron collider 
were carried out at the Fermilab Tevatron, assuming the 
production channel
$q\overline q\to H^{++}H^{--}$ and decay $H^{\pm\pm}\to \ell^\pm_i\ell^\pm_j$. 
The mass limits $M_{H^{\pm\pm}}> 110\to 150$ GeV \cite{Acosta:2004uj,Abazov:2004au,:2008iy,Aaltonen:2008ip} were derived, with the
strongest limits being for $\ell=e,\mu$ \cite{Acosta:2004uj,Abazov:2004au,:2008iy}.
The branching ratios (BRs) for 
$H^{\pm\pm}\to \ell^\pm_i\ell^\pm_j$ depend on $h_{ij}$
and are predicted in the HTM in terms of the parameters
of the neutrino mass matrix
\cite{Akeroyd:2005gt, Ma:2000wp, Chun:2003ej}.
Detailed quantitative studies of
BR($H^{\pm\pm}\to \ell^\pm_i\ell^\pm_j$) (and BR($H^{\pm}\to \ell^\pm_i\nu$))
in the HTM have been performed in
\cite{Garayoa:2007fw,Akeroyd:2007zv,Kadastik:2007yd,Perez:2008ha,delAguila:2008cj,Akeroyd:2009hb} 
with particular emphasis  
given to their sensitivity to the Majorana phases and 
the absolute neutrino mass i.e. parameters which cannot be 
probed in neutrino oscillation experiments. 
Simulations of the detection prospects of $H^{\pm\pm}$ at the LHC 
with $\sqrt s=14$ TeV previously focussed on $q\overline q\to \gamma^*,Z^* \to H^{++}H^{--}$ 
only \cite{Azuelos:2005uc, Rommerskirchen:2007jv},
but recent studies now include
the mechanism $q'\overline q\to H^{\pm\pm}H^{\mp}$ \cite{Perez:2008ha, delAguila:2008cj, Akeroyd:2010ip}. 
The first search
for $H^{\pm\pm}$ at the LHC with $\sqrt s=7$ TeV \cite{CMS-search}
has recently been performed for both
production mechanisms
$q\overline q\to H^{++}H^{--}$ and $q\overline {q'}\to H^{\pm\pm}H^{\mp}$,
for the decay channels $H^{\pm\pm}\to \ell^\pm_i\ell^\pm_j$ and $H^{\pm}\to \ell^\pm_i\nu$
(where $i,j=e,\mu,\tau$).

In the HTM there are two electrically neutral mass eigenstates which are CP-even scalars. These are
denoted by $H_1$ and $H_2$, with $M_{H_1} < M_{H_2}$. One of the eigenstates is dominantly composed of
the isospin doublet field (and plays the role of the SM Higgs boson) while the other eigenstate 
is dominantly composed of the real part of the neutral triplet field. The mixing angle is small because
it depends on the small ratio $v_\Delta/v < 0.03$ (where $v=246$ GeV, the vev of the doublet field).
In phenomenological studies of the HTM it is usually assumed that the lighter eigenstate $H_1$ is 
the one which is dominantly composed of the isospin doublet field. Therefore the phenomenology of $H_1$ is 
more or less identical to that of the SM-Higgs boson. The converse case
of the heavier eigenstate $H_2$ being the one which is dominantly composed of the isospin doublet field is possible in the 
HTM, and has been mentioned in \cite{Dey:2008jm, Akeroyd:2010je,Arhrib:2011uy}.
However, no detailed study of the phenomenology of $H_2$ in such a scenario has been carried out.
Importantly, if $M_{H_2}> 2M_{\phi}$ \cite{Arhrib:2011uy}
(where $\phi$ is one of the dominantly triplet eigenstates $H_1, A^0, H^\pm, H^{\pm\pm}$)
then new decay channels for $H_2$ would be possible. This would give rise to a phenomenology 
of $H_2$ which differs somewhat from that of the SM Higgs boson.
In this work we focus on the case of $M_{H_2}> 2M_{H^{\pm\pm}}$, because 
a new production mechanism for $H^{\pm\pm}$ would be possible, namely gluon-gluon fusion $gg\to H_2$ followed by
decay $H_2\to H^{++}H^{--}$. The case of $M_{H_2}> 2M_{H^{\pm\pm}}$ necessarily requires $M_{H_2}\gsim 200$ GeV,
a mass region which is now being probed for the first time 
by the LHC for the decay channels to SM particles, $H_2\to WW$ and $H_2\to ZZ$ \cite{Collaboration:2011qi,Chatrchyan:2011tz}.

Our work is organised as follows. In section II we describe the theoretical structure of the HTM. In section III the
parameter space for $M_{H_2}> 2M_{\phi}$ (where $\phi$ is one of $H_1, A^0, H^\pm, H^{\pm\pm}$) is described.
In section IV the formulae for the decay widths of $H_2\to H^{++}H^{--},H^{+}H^{-},A^0A^0,H_1H_1$ are presented. 
Section V contains a numerical analysis of the magnitude of the branching ratios of the above channels, 
as well as a quantitative study of the cross section for pair production of $H^{\pm\pm}$
via $gg\to H_2$, with decay $H_2\to H^{++}H^{--}$. Conclusions are given in section VI.

\section{The Higgs Triplet Model}
In the HTM~\cite{Konetschny:1977bn,Schechter:1980gr,Cheng:1980qt}
a $I=1,Y=2$ complex $SU(2)_L$ isospin triplet of 
scalar fields is added to the SM Lagrangian. 
Such a model can provide a Majorana mass for the observed neutrinos 
without the introduction of a right-handed neutrino via the 
gauge invariant Yukawa interaction:
\begin{equation}
{\cal L}=h_{ij}\psi_{iL}^TCi\tau_2\Delta\psi_{jL}+h.c
\label{trip_yuk}
\end{equation}
Here $h_{ij} (i,j=e,\mu,\tau)$ is a complex
and symmetric coupling,
$C$ is the Dirac charge conjugation operator, $\tau_2$
is a Pauli matrix,
$\psi_{iL}=(\nu_i, \ell_i)_L^T$ is a left-handed lepton doublet,
and $\Delta$ is a $2\times 2$ representation of the $Y=2$
complex triplet fields:
\begin{equation}
\Delta
=\bma{cc}
\delta^+/\sqrt{2}  & \delta^{++} \\
\delta^0       & -\delta^+/\sqrt{2}
\ema
\end{equation}
A non-zero triplet vacuum expectation value $\langle\Delta^0\rangle$ 
gives rise to the following mass matrix for neutrinos:
\begin{equation}
m_{ij}=2h_{ij}\langle\Delta^0\rangle = \sqrt{2}h_{ij}v_{\Delta}
\label{nu_mass}
\end{equation}
The necessary non-zero $v_{\Delta}$ arises from the minimisation of
the most general $SU(2)\otimes U(1)_Y$ invariant Higgs potential,
which is written as follows~\cite{Ma:2000wp, Chun:2003ej}
(with $\Phi=(\phi^+,\phi^0)^T$):

\begin{eqnarray}
V(H,\Delta) & = & 
- m_H^2 \ H^\dagger H \ + \ \lambda (H^\dagger H)^2 \ 
+ \ M_{\Delta}^2 \ {\rm Tr} \Delta^\dagger \Delta\ 
+ \ \left( \mu \ H^T \ i \sigma_2 \ \Delta^\dagger H \ + \ {\rm h.c.}\right) \ 
\nonumber \\
&& + \ \lambda_1 \ (H^\dagger H) {\rm Tr} \Delta^\dagger \Delta \ 
+ \ \lambda_2 \ \left( {\rm Tr} \Delta^\dagger \Delta \right)^2 \ 
+ \ \lambda_3 \ {\rm Tr} \left( \Delta^\dagger \Delta \right)^2 \ 
+ \ \lambda_4 \ H^\dagger \Delta \Delta^\dagger H 
\label{Potential}
\end{eqnarray}

Here $m^2_H<0$ in order to ensure $\langle\phi^0\rangle=v/\sqrt 2$ which
spontaneously breaks $SU(2)\otimes U(1)_Y$
to  $U(1)_Q$, and $M^2_\Delta\,(>0)$ is the mass term for the triplet scalars.
In the model of Gelmini-Roncadelli~\cite{Gelmini:1980re} 
the term $\mu(\Phi^Ti\tau_2\Delta^\dagger\Phi)$ is absent,
which leads to spontaneous violation of lepton number for $M^2_\Delta<0$.
The resulting Higgs
spectrum contains a massless triplet scalar (a Majoron, $J$) and another light 
scalar ($H^0$). Pair production via $e^+e^-\to H^0J$ would give a large 
contribution to the invisible width of the $Z$ boson and this model
was excluded at the CERN Large Electron Positron Collider (LEP). 
The inclusion of the term $\mu(\Phi^Ti\tau_2\Delta^\dagger\Phi$)
explicitly breaks lepton number when $\Delta$ is assigned $L=2$, and eliminates
the Majoron~\cite{Konetschny:1977bn,Schechter:1980gr,Cheng:1980qt}.
Thus the scalar potential in
eq.~(\ref{Potential}) together with the triplet Yukawa interaction of
eq.~(\ref{trip_yuk}) lead to a phenomenologically viable model of neutrino mass
generation.
 For small $v_\Delta/v$,
the expression for $v_\Delta$
resulting from the minimisation of $V$ is:
\begin{equation}
v_\Delta \simeq \frac{\mu v^2}{\sqrt 2 (M^2_\Delta+v^2(\lambda_1+\lambda_4)/2)} 
\label{tripletvev}
\end{equation}
For large $M_\Delta$ compared to $v$
one has $v_\Delta \simeq \mu v^2/\sqrt 2 M^2_\Delta$,
which is sometimes referred to as the ``Type II seesaw mechanism''
and would naturally lead to a small $v_\Delta$.
Recently there has been much interest in 
the scenario of light triplet scalars ($M\approx v$),  (especially the distinctive doubly charged scalar, $H^{\pm\pm}$),
within the discovery reach of
the LHC, for which eq.~(\ref{tripletvev}) leads to $v_\Delta\approx \mu$.
In extensions of the HTM the term $\mu(\Phi^Ti\tau_2\Delta^\dagger\Phi$) 
may arise in various ways: i) it can be generated at tree level via the vev of a Higgs singlet field%
~\cite{Schechter:1981cv}; 
ii) it can arise at higher orders in perturbation theory~\cite{Chun:2003ej};
iii) it can originate in the context of extra dimensions \cite{Ma:2000wp};
iv) it can arise in models with an additional heavy scalar triplet \cite{Majee:2010ar}.
 
Some phenomenological studies focus on a simplified scalar 
potential (e.g. Ref.~\cite{Perez:2008ha}) in which the 
quartic couplings $\lambda_i$ (where $i=1,2,3,4$) involving the triplet field $\Delta$
are neglected. The resulting scalar potential then depends on four parameters
 ($-m^2_H$, $\lambda$, $\mu$, $M_{\Delta}$), but only three parameters are 
independent because the VEV for the doublet field ($v=246$ GeV) is fixed by the
 mass of $W^\pm$. The three independent parameters are usually chosen as 
$\lambda,v_\Delta,M_{\Delta}$ or $\lambda,v_\Delta,\mu$
(see eq.~(\ref{tripletvev})).
 The inclusion of $\lambda_i$ generates additional trilinear and quartic 
couplings among the scalar mass eigenstates.  
The terms with $\lambda_1$ and $\lambda_4$, which involve both triplet and 
doublet fields, are of particular interest because they can give 
a sizeable contribution to the masses of the scalar eigenstates (see below).
A detailed study of the theoretical constraints on the scalar potential
(e.g. vacuum stability, unitarity and perturbativity) 
has been carried out in \cite{Arhrib:2011uy}.

An upper limit on $v_\Delta$ can be obtained from
considering its effect on the parameter $\rho (=M^2_W/M_Z^2\cos^2\theta_W)$. 
In the SM $\rho=1$ at tree-level, while in the HTM one has
(where $x=v_\Delta/v$):
\begin{equation}
\rho\equiv 1+\delta\rho={1+2x^2\over 1+4x^2}
\label{deltarho}
\end{equation}
The measurement $\rho\approx 1$ leads to the bound
$v_\Delta/v\lsim 0.03$, or  $v_\Delta\lsim 8$~GeV\@.
Production mechanisms which depend on $v_\Delta$ 
(i.e. $pp\to W^{\pm *}\to W^\mp H^{\pm\pm}$ and fusion via $W^{\pm *} W^{\pm *} \to H^{\pm\pm}$
\cite{Huitu:1996su,Gunion:1989ci}) are not competitive with the processes 
$q\overline q\to H^{++}H^{--}$ and $\overline q q'\to H^{\pm\pm}H^{\mp}$
at the energies of the Fermilab Tevatron, but such mechanisms could be the dominant source of 
$H^{\pm\pm}$ at the LHC if $v_{\Delta}={\cal O}$ (1 GeV) and 
$M_{H^{\pm\pm}}> 500$ GeV.
At the 1-loop level, $v_\Delta$ must be renormalised and explicit
analyses lead to bounds on its magnitude similar to the above bound from
the tree-level analysis, e.g.\ see \cite{Blank:1997qa,Chen:2006pb}.

The scalar eigenstates in the HTM are as follows: i) the charged scalars
 $H^{\pm\pm}$ and $H^\pm$; ii) the CP-even neutral scalars
$H_1$ and $H_2$; iii) a CP-odd neutral scalar $A^0$.
The doubly charged $H^{\pm\pm}$ is entirely composed of the triplet 
scalar field $\Delta^{\pm\pm}$, 
while the remaining eigenstates are in general mixtures of the  
doublet and triplet fields. However, such mixing is proportional to the 
triplet vev, and hence small {\it even if} $v_\Delta$
assumes its largest value of a few GeV\@.
\footnote{A large mixing angle is possible in the CP-even sector
provided that $M_{H_1}\sim M_{H_2}$ \cite{Akeroyd:2010je,Dey:2008jm,Arhrib:2011uy}.}

\section{Scenario of a heavy SM-like Higgs boson $(H_2)$ in the HTM}

In the HTM there are two CP-even mass eigenstates, 
which we denote by $H_1$ and $H_2$ (where both $M_{H_2}>M_{H_1}$
and $M_{H_1}>M_{H_2}$ are possible). Their compositions in terms of 
the original fields of the Lagrangian are as follows: 
\begin{eqnarray}
H_2 &=& \cos \theta_0 \ h^0 \ + \ \sin \theta_0 \ \Delta^0
\quad
H_1 = - \sin \theta_0 \ h^0 \ + \ \cos \theta_0 \Delta^0
\qquad
\label{CP-even-mix}
\end{eqnarray}
Here $h^0$ is the real part of the electrically neutral doublet field $\phi^0$, and 
$\Delta^0$ is the real part of the electrically neutral triplet field $\delta^0$. 
The mixing angle $\theta_0$ is very small,\footnote{The mixing angle
can be maximal in the region of degeneracy $M_{H_2}\sim M_{H_1}$, but it quickly becomes small ($\sim v_\Delta/v$) with increasing mass splitting
$|M_{H_2}-M_{H_1}|$\cite{Akeroyd:2010je,Dey:2008jm,Arhrib:2011uy}.}
 being of order 0.03 at most ($\sin \theta_0\sim v_\Delta/v$).
Hence $H_2$ is essentially composed of the doublet field $h^0$, with couplings to the fermions and gauge bosons
which are almost identical 
to those of the SM Higgs boson, while $H_1$ is mainly composed of the triplet field $\Delta^0$.

The explicit expression for the $2\times 2$ CP-even scalar mass matrix for the
scalar potential in eq.~(\ref{Potential}) is given in 
several works e.g. \cite{Akeroyd:2010je,Dey:2008jm,Arhrib:2011uy}.
Neglecting the small off-diagonal elements in this mass matrix, the approximate expressions for the squared masses of $H_1$ and $H_2$ are
as follows:
\begin{eqnarray}
M^2_{H_2}= 
2\lambda v^2  \\ 
\label{massH2}
M^2_{H_1}= M^2_\Delta + (\frac{\lambda_1}{2} 
 +\frac{\lambda_4}{2}) v^2 + 3(\lambda_2 +\lambda_3) v^2_\Delta
\label{massH1}
\end{eqnarray}
The squared mass of the (dominantly triplet) CP-odd $A^0$ is given by:
\begin{equation}
M^2_{A^0} = M^2_\Delta + (\frac{\lambda_1}{2} 
 +\frac{\lambda_4}{2}) v^2 + (\lambda_2 +\lambda_3) v^2_\Delta
\label{CP-odd-mass}
\end{equation}
The squared mass of the (dominantly triplet) $H^\pm$ is given by:
\begin{equation}
M^2_{H^\pm} = M^2_\Delta + (\frac{\lambda_1}{2} 
 +\frac{\lambda_4}{4}) v^2 + (\lambda_2  + \sqrt 2
 \lambda_3) v^2_\Delta
 \qquad
\label{charged-mass}
\end{equation}
Finally, the squared mass of the (purely triplet) doubly-charged 
scalar ($H^{\pm\pm} = \delta^{\pm\pm}$) is given by:
\begin{equation}
M^2_{H^{\pm\pm}} =M^2_\Delta + \frac{\lambda_1}{2} v^2
+\lambda_2 v_\Delta^2 
\label{doub-charged-mass}
\end{equation}
One can see that the squared mass of the (dominantly doublet) $H_2$
 is simply given by $2\lambda v^2$, as in the
SM. In the expressions for the masses
of $M^2_{A^0}$, $M^2_{H_1}$, $M^2_{H^\pm}$ and $M^2_{H^{\pm\pm}}$ 
there is a common term $M^2_\Delta + \frac{\lambda_1}{2} v^2$.
It is evident that the mass scales for $H_2$ and
the dominantly triplet scalars ($A^0,H_1, H^\pm, H^{\pm\pm}$) are unrelated, the former being set 
by $2\lambda v^2$ and the latter by $M^2_\Delta + \frac{\lambda_1}{2} v^2$.
Neglecting the terms which are proportional to the small parameter $v_\Delta$, one can see that 
there are only two possible mass hierarchies for the triplet scalars, with the magnitude of the
mass splitting being controlled by $\lambda_4$ (and $M_{A^0}=M_{H_1}$):
\begin{eqnarray}
M_{A^0},M_{H_1} < M_{H^\pm} < M_{H^{\pm\pm}} \;\;{\rm for}\;\; \lambda_4<0
\label{mass-split1} \\
M_{H^{\pm\pm}}  < M_{H^\pm} <  M_{A^0},M_{H_1} \;\; {\rm for} \;\; \lambda_4>0
\label{mass-split2}
\end{eqnarray}

In studies of the HTM it is sometimes assumed that $M^2_\Delta \gg  2\lambda v^2$
i.e. $M_{H^{\pm\pm}}, M_{H^\pm},M_{H_1},M_{A^0} \gg M_{H_2}$. The motivation for this scenario is 
to have a ``seesaw type'' explanation for the smallness of $v_\Delta$ in eq.~(\ref{tripletvev}).
However, for $M_\Delta$ much larger than the TeV scale there would be no hope of observing the triplet scalars
at the LHC. In recent years there has been much interest in the study of the HTM as a TeV scale model of neutrino mass generation \cite{Muhlleitner:2003me,Han:2007bk,Huitu:1996su,Dion:1998pw,Akeroyd:2005gt,Ma:2000wp,Chun:2003ej,Garayoa:2007fw,Akeroyd:2007zv,Kadastik:2007yd,Perez:2008ha,delAguila:2008cj,Akeroyd:2009hb,Azuelos:2005uc,Rommerskirchen:2007jv,Akeroyd:2010ip}
i.e. not invoking a large mass scale for $M_\Delta$. In these studies it is assumed
(either explicitly or implicitly) that $M^2_\Delta >  2\lambda v^2$, with $M_\Delta < 1$ TeV. 

The converse case where $M^2_\Delta + \frac{\lambda_1}{2} v^2<  2\lambda v^2$ 
is rarely considered.
In \cite{Akeroyd:2010je,Dey:2008jm,Arhrib:2011uy} the possibility of
$M_{H_2} >  M_{H^{\pm\pm}}, M_{H^\pm},M_{H_1}, M_{A^0}$ has been mentioned, and
in \cite{Arhrib:2011uy} the case of 
$M_{H_2}> 2M_{\phi}$ (where $\phi$ is one of $H^{\pm\pm},H^\pm, H_1, A^0$)
is explicitly discussed. However, in these works there is no study of the phenomenology of 
$H_2$ at hadron colliders for the case of $M_{H_2}> 2M_{\phi}$, and how its
experimental signature might differ from that of the SM Higgs Boson.
Importantly, if $M_{H_2}> 2M_{\phi}$ then
new decay channels 
for $H_2$ become possible,\footnote{The scenario of a heavy SM-like Higgs boson decaying to
singly charged scalars, $h^0\to H^+H^-$, has been discussed in the Two Higgs Doublet Model \cite{deVisscher:2009zb}.} 
namely $H_2\to H^{++}H^{--},H^{+}H^{-},H_1H_1,A^0A^0$.

In this case the phenomenology of $H_2$ in the HTM could be different to that of the SM Higgs boson,
because the new decay channels (if open kinematically) would compete with the usual decays 
of $H_2$ to SM particles 
(i.e. $H_2\to WW,ZZ,t\overline t$).  
Of particular interest is the decay $H_2\to H^{++}H^{--}$, for which the
condition $M_{H_2}> 2M_{H^{\pm\pm}}$ is necessary. If its branching ratio were sizeable then
the production of $H_2$ via gluon-gluon fusion $gg \to H_2$ followed by the decay $H_2\to H^{++}H^{--}$
would be an additional way to produce a pair of $H^{\pm\pm}$ at hadron colliders.

We note that the condition $M_{H_2}> 2M_{H^{\pm\pm}}$ necessarily requires 
$M_{H_2}\gsim 200$ GeV in order to respect the current lower bounds on $M_{H^{\pm\pm}}$
from direct searches. At first sight, such a heavy SM-like $H_2$ would appear to be 
in conflict with experimental data,  since it is well known that the Higgs boson in the SM is expected to be lighter than 200 GeV
in order not to give an unacceptably large contribution to electroweak precision observables.
In the context of the SM the case of $M_{H_2}\gsim 200$ GeV is quite strongly disfavoured, although 
this fact has not dissuaded direct searches in this mass region at the LHC \cite{Collaboration:2011qi,Chatrchyan:2011tz}.
However, the bound  $M_{H_2}\gsim 200$ GeV cannot strictly be applied to the
HTM, due to the additional scalar particles and the different renormalisation procedure,
the latter being necessary because of the presence of the triplet vev ($v_\Delta$). 
Dedicated analyses in models with scalar triplets have shown that a heavy (up to 1 TeV) SM-like
Higgs boson can be made consistent with electroweak precision measurements \cite{Blank:1997qa,Chen:2006pb}. 
These studies are for a model with a real $Y=0$ scalar triplet, which has no doubly charged scalar and
gives $\rho>1$ at tree level, in contrast to the HTM which has $\rho<1$ at tree level, (eq.~(\ref{deltarho})).
One can see in \cite{Chen:2006pb}
that the condition $M_{H_2}> 2M_{\phi}$ (where $\phi$ is one of the $Y=0$ triplet scalars) can be
be accommodated. Although there is no explicit study in the HTM, we expect this result to also hold due to 
its greater number of free parameters (i.e. particle masses).
In our numerical analysis we will treat $M_{H_2}< 700$ GeV and
$M_{H_2}> 2M_{H^{\pm\pm}}$ as permissible parameter space in the HTM.

From a phenomenological point of view, a heavy ($>> 200$ GeV) SM-like Higgs boson
is attractive because it would be discovered more 
quickly at the LHC than a light SM-like Higgs boson with mass $< 140$ GeV. The region of 
$200 \; {\rm GeV} < M_{H_2}  <  500 \; {\rm GeV}$, for which the decays $H_2\to ZZ$ and $H_2 \to WW$ are dominant in the SM,
is a mass range where
the LHC has sensitivity to cross sections which are much smaller than that of the SM Higgs boson. 
The first searches at the LHC for a SM Higgs with $M_{H_2}>200$ GeV
have already been carried out. The ATLAS collaboration
(with 36 pb$^{-1}$ of integrated luminosity) has searched for $H_2\to ZZ$ with the decay modes
$ZZ\to \ell^+\ell^-\nu\nu, ZZ\to \ell^+\ell^-q\overline q$ and $ZZ\to \ell^+\ell^-\ell^+\ell^-$,
as well as $H_2\to WW$ with the decay mode $WW\to \ell\nu q'\overline q$ \cite{Collaboration:2011qi}.
The CMS collboration has searched for  $H_2\to WW$ with the decay mode $WW\to \ell\nu\ell\nu$ \cite{Chatrchyan:2011tz}.
Production of $H_2$ is assumed to be via gluon-gluon fusion, $gg\to H_2$, and
cross sections which are an order of magnitude above the prediction of the SM 
are currently being excluded at $95\%$ c.l. By the end of the $\sqrt s=7$ TeV run
(in which a few fb$^{-1}$ of integrated luminosity will be accumulated), the
sensitivity in these channels will be sufficient to exclude or provide evidence for the SM Higgs boson
at a high confidence level in the region $200 \; {\rm GeV} < M_{H_2}  <  500 \; {\rm GeV}$.
If the branching ratios of $H_2\to H^{++}H^{--},H^{+}H^{-},H_1H_1,A^0A^0$ were sizeable then discovery of $H_2$ in the channels 
$H_2\to ZZ$ and $H_2\to WW$ would require more integrated luminosity.

\section{The decays $H_2\to \phi\phi$ with $\phi=H^{\pm\pm}, H^\pm,H_1, A^0$}

There are four decay channels of $H_2$ to pairs of scalars in the HTM:
$H_2\to H^{++}H^{--}$, $H_2\to H^{+}H^{-}$, $H_2\to A^0A^0$ and 
$H_2\to H_1H_1$. If $M_{H_2}>2M_\phi$ (where 
$\phi=H^{\pm\pm}, H^\pm,H_1, A^0$) one can treat this as a two-body decay to a pair of on-shell $\phi$.
If $M_{H_2}< 2M_\phi$ we consider the partial width to be zero.
Between one and four of the decays $H_2\to \phi\phi$ can be open kinematically,
depending on the mass splitting among $\phi$ (which is controlled by $\lambda_4$ in eq.~(\ref{mass-split1}) and 
eq.~(\ref{mass-split2}).
For $\lambda_4>0$ the lightest of the triplet scalars is $H^{\pm\pm}$.
In this scenario, once values of $M_{H_2}$ and $M_{H^{\pm\pm}}$ are chosen such that $M_{H_2} > 2M_{H^{\pm\pm}}$ 
there will be a value of $\lambda_4$ above which only $H_2\to H^{++}H^{--}$ is kinematically open. This
will be the scenario where BR($H_2\to H^{++}H^{--})$ is maximal.

The Feynman rules for the scalar trilinear couplings which mediate the decays
are as follows (omitting a factor of $-i$):
\begin{eqnarray}
C_{H_2H^{++}H^{--}}= \lambda_1v 
\label{doub-coup} \\
C_{H_2H^{+}H^{-}}= (\lambda_1+\frac{\lambda_4}{2})v 
\label{sing-coup} \\
C_{H_2H_1H_1},C_{H_2A^0A^0} = (\lambda_1+\lambda_4)v
\label{neut-coup}
\end{eqnarray}
Here we consider $H_2$ to be entirely composed of the isospin doublet scalar field, which is true to a very good approximation.
One can see that $C_{H_2H^{++}H^{--}}$ is controlled only by
$\lambda_1$, while the other trilinear couplings depend on both $\lambda_1$ and $\lambda_4$.
If $\lambda_1$ and $\lambda_4$ are sizeable, then the branching ratios for 
$H_2\to H^{++}H^{--}$, $H_2\to H^{+}H^{-}$, $H_2\to H_1H_1$, and $H_2\to A^0A^0$ could be non-negligible.

One can use a generic formula for the decay rate for the four channels:
\begin{equation}
\Gamma(H_2\to \phi\phi)=\delta_H\frac{|C_{H_2\phi \phi}|^2}{32\pi M_{H_2}}\left
(1-\frac{4M^2_{\phi}}{M^2_{H_2}}\right)^{1/2}
\label{scalar-width}
\end{equation}
Here $\delta_H=2$ for
$\phi=H^{\pm\pm}, H^\pm$ (i.e. non-identical particles in the final state)
and $\delta=1$ for $\phi=H_1, A^0$ (i.e. identical particles in the final state)

It is clear that the two crucial parameters for a large BR($H_2\to H^{++}H^{--}$)
are $\lambda_1$ (which determines the strength of the coupling
$|C_{H_2H^{++}H^{--}}|$ in eq.~(\ref{doub-coup}) and $M_{H^{\pm\pm}}$ (which determines the suppression from phase space).
In our numerical analysis we shall take $M_{H^{\pm\pm}}$ as an input parameter. As can be seen from 
eq.~(\ref{doub-charged-mass}), the dominant contribution to $M^2_{H^{\pm\pm}}$ is
from the two terms $M^2_\Delta+\lambda_1v^2/2$. Therefore, by taking $\lambda_1$
and $M_{H^{\pm\pm}}$ as input parameters the value of $M^2_\Delta$ is determined. We will be focussing on the parameter space of 
$90 \,{\rm GeV} < M_{H^{\pm\pm}} < 300 \,{\rm GeV}$
and $0 < \lambda_1 < 4$, and consequently $M^2_\Delta<0$ 
when $M^2_{H^{\pm\pm}}<\lambda_1v^2/2$. In the scenario of $M^2_\Delta<0$ the positive mass of
$M_{H^{\pm\pm}}$ is obtained from the term $\lambda_1v^2/2$.
Alternatively, one could consider
$\lambda_1<0$ and $M^2_\Delta>0$. The crucial point here is that 
the parameters $M^2_\Delta$ and $\lambda_1$ should have opposite signs if one
wishes to have large $|\lambda_1|$ (in order to enhance $|C_{H_2H^{++}H^{--}}|$) 
together with a fairly light $H^{\pm\pm}$.

We now summarise the current lower limits on  $M_{H^{\pm\pm}}$ from direct searches.
There have been searches for the decay channels $H^{\pm\pm}\to \ell^\pm_i\ell^\pm_j$ for $i,j=e,\mu,\tau$ (these are the dominant decay
channels for $v_\Delta \lsim 0.1$ MeV) at LEP \cite{Abdallah:2002qj}, 
Tevatron \cite{Acosta:2004uj,Abazov:2004au,:2008iy,Aaltonen:2008ip}
and the LHC \cite{CMS-search} (CMS Collaboration). The strongest mass limits for the decays $H^{\pm\pm}\to ee,e\mu,\mu\mu$ are
from the LHC search, which obtained $M_{H^{\pm\pm}}> 144, 154, 156$ GeV respectively, assuming BR$=100\%$ in a given channel.
Separate searches for three and four leptons (which have significantly different backgrounds) were performed. These 
limits are weakened considerably for the case of BR$<100\%$ because the event number for the signal is proportional to
the square of the branching ratio (${\rm BR}^2$). In \cite{CMS-search} both production mechanisms 
$q\overline q\to \gamma^*,Z^* \to H^{++}H^{--}$ and $q'\overline q \to W \to H^{\pm\pm}H^{\mp}$ were
considered (with the assumption $M_{H^{\pm\pm}}=M_{H^{\pm}}$), which increases the sensitivity in the
three-lepton channel. For the decays involving one $\tau$, namely $H^{\pm\pm}\to e\tau,\mu\tau$,
the limit $M_{H^{\pm\pm}}> 106$ GeV was derived in both channels at the LHC \cite{CMS-search}, with stronger limits  
from the Tevatron ($M_{H^{\pm\pm}}> 112$ GeV and 114 GeV respectively) obtained in \cite{Aaltonen:2008ip}.
The only search for $H^{\pm\pm}\to \tau^\pm\tau^\pm$ at a hadron collider is the LHC search in \cite{CMS-search}, which 
derived the limit $M_{H^{\pm\pm}} \gsim 80$ GeV.

We will take $M_{H^{\pm\pm}}=90$ GeV as our lowest value for the mass of $H^{\pm\pm}$, and this is allowed for certain choices
of branching ratios of $H^{\pm\pm}$. As explained above, the limits on $M_{H^{\pm\pm}}$ from hadron colliders are weakest for those
channels which involve $\tau$. In contrast, the limit from the LEP searches of $M_{H^{\pm\pm}} \gsim 100$ GeV applies to all the decays 
$H^{\pm\pm}\to \ell^\pm_i\ell^\pm_j$ with $i,j=e,\mu,\tau$.
The search strategy at LEP requires four leptons and so the event number for the signal is proportional to ${\rm BR}^2$. 
The scenario of $M_{H^{\pm\pm}}=90$ GeV is compatible with the all the above searches
provided that the decays involving $\tau$ are dominant e.g. choices like 
BR$(H^{\pm\pm}\to e\tau,\mu\tau,\tau\tau)$ of around 33\%. 
It is not necessary to have BR$(H^{\pm\pm}\to ee,e\mu,\mu\mu)$ 
totally absent for $M_{H^{\pm\pm}}=90$ GeV, and 
BRs of the order of $10\%$ for these channels can be accommodated because the event number is proportional to
${\rm BR}^2$, and for ${\rm BR}=10\%$ this is a large suppression factor.
We note that the sum of BR$(H^{\pm\pm}\to ee,e\mu,\mu\mu)$ cannot be taken arbitrarily small in the HTM because
the Yukawa couplings $h_{ij}$ are related to the neutrino mass matrix via eq.~(\ref{nu_mass}). The allowed values of 
BR$(H^{\pm\pm}\to \ell^\pm_i\ell^\pm_j)$ in the HTM have been studied in detail in 
\cite{Garayoa:2007fw,Akeroyd:2007zv,Kadastik:2007yd,Perez:2008ha,delAguila:2008cj,Akeroyd:2009hb}, and in \cite{delAguila:2008cj}
it can be seen explicitly that the sum of BR$(H^{\pm\pm}\to ee,e\mu,\mu\mu)$ must be greater than around 5\%.

Very recently the searches for $H^{\pm\pm}\to ee,e\mu,\mu\mu$ by the CDF collaboration in \cite{Acosta:2004uj}
(which used 0.24 fb$^{-1}$) were updated using 6.1 fb$^{-1}$ \cite{Aaltonen:2011rt}. Mass limits of 
$M_{H^{\pm\pm}}> 225, 210, 245$ GeV were obtained, again assuming BR$=100\%$. In these searches the event number
for the signal is linear in BR, and for BR$\sim 3\%$($15\%$) the limit $M_{H^{\pm\pm}}> 245$ GeV for $H^{\pm\pm}\to \mu\mu$ would weaken to
 $M_{H^{\pm\pm}}> 100$ GeV (150 GeV). Note that these mass limits in \cite{Aaltonen:2011rt}
 only assume production of $H^{\pm\pm}$ from 
$q\overline q\to \gamma^*,Z^* \to H^{++}H^{--}$. The inclusion of
$q'\overline q \to W \to H^{\pm\pm}H^{\mp}$ would allow larger values of $M_{H^{\pm\pm}}$ to be probed.
Finally, if the decay $H^{\pm\pm}\to W^\pm W^\pm$ is dominant (which is the case for
$v_\Delta \gsim 0.1$ MeV) then $M_{H^{\pm\pm}}=90$ GeV is
permitted because there have been no direct searches for this channel in the context
of models with $H^{\pm\pm}$. We will respect the all the above mass limits in our numerical analysis, the most
stringent ones being for the channels $H^{\pm\pm}\to e^\pm e^\pm, e^\pm \mu^\pm$ and $\mu^\pm\mu^\pm$.

We will only consider the scenario of $M_{H_2}>200$ GeV for which the decay channels $H_2\to WW$ and 
$H_2\to ZZ$ can be treated as two-body decays. The expressions for their partial decay widths are as follows:
\begin{eqnarray}
\Gamma(H_2 \to WW) =  \frac{\sqrt{2}G_F M^3_{H_2}}{32 \pi}(1-4\kappa_W+12\kappa_W^2)
(1-4\kappa_W)^{1/2}  \delta_W
\label{WWwidth}
\end{eqnarray}
with $\kappa_W=M_W^2/M_{H_2}^2$ and $\delta_W=2$.
\begin{eqnarray}
\Gamma(H_2 \to ZZ) =  \frac{\sqrt{2}G_F M^3_{H_2}}{32 \pi}(1-4\kappa_Z+12\kappa_Z^2)
(1-4\kappa_Z)^{1/2}  \delta_Z
\label{ZZwidth}
\end{eqnarray}
with $\kappa_Z=M_Z^2/M_{H_2}^2$ and $\delta_Z=1$.
If $M_{H_2}> 2m_t$ then the decay channel $H_2\to t\overline t$ is open:
\begin{eqnarray}
\Gamma(H \to \bar{t} t) & = &  \frac{3 G_F m_t^2}{4\sqrt{2} \pi}
 \ M_{H}  \beta^{3}_t
\label{ttwidth}
\end{eqnarray}
where $\beta_t=(1- 4m_t^2/M_{H_2}^2)^{1/2}$.
All other decays of $H_2$ to SM particles (e.g. $H_2\to b\overline b, \tau^+\tau^-$) 
have negligibly small partial widths for $M_{H_2}>200$ GeV. Note that other decay channels
such as $H_2\to H^\pm W$, $H_2\to H_1 W$ and $H_2\to A^0 Z$ are suppressed by 
the small mixing between the doublet and triplet fields, and so can be neglected.

\section{Numerical Analysis} 

We now study the magnitude of the branching ratios of the decays channels
$H_2\to \phi\phi$ for $\phi=H^{\pm\pm}, H^\pm,H_1, A^0$. The four important
parameters are $M_{H^{\pm\pm}}$, $M_{H_2}$, $\lambda_1$ and
$\lambda_4$. The other parameters in the scalar potential are fixed as 
$\lambda_2=\lambda_3=0.5$ and $v_\Delta=10^{-2}$ MeV, the latter choice ensuring that
the decays $H^{\pm\pm}\to \ell^\pm\ell^\pm$ are dominant. These latter three parameters
appear in the expressions for the masses of the triplet scalars in
eq.~(\ref{massH1}) to eq.~(\ref{doub-charged-mass})  
but their effect is essentially negligible, even for the case of  $v_\Delta=1$ GeV.
We will present results for $M_{H^{\pm\pm}}=90$ GeV, 150 GeV, 200 GeV and 300 GeV.
As explained in the previous section, the choice of $M_{H^{\pm\pm}}=90$ GeV requires small BRs ($< 3\%$) for the decay channels
$H^{\pm\pm}\to e^\pm e^\pm, e^\pm \mu^\pm$ and $\mu^\pm\mu^\pm$ (and consequently large BRs to channels
involving $\tau$) in order to respect the limits from the direct searches for $H^{\pm\pm}$.
Larger values ($\gg 3\%$) of BR($H^{\pm\pm}\to e^\pm e^\pm, e^\pm \mu^\pm, \mu^\pm\mu^\pm$) are permitted as $M_{H^{\pm\pm}}$ increases.

\begin{figure}[t]
\begin{center}
\includegraphics[origin=c, angle=0, scale=0.46]{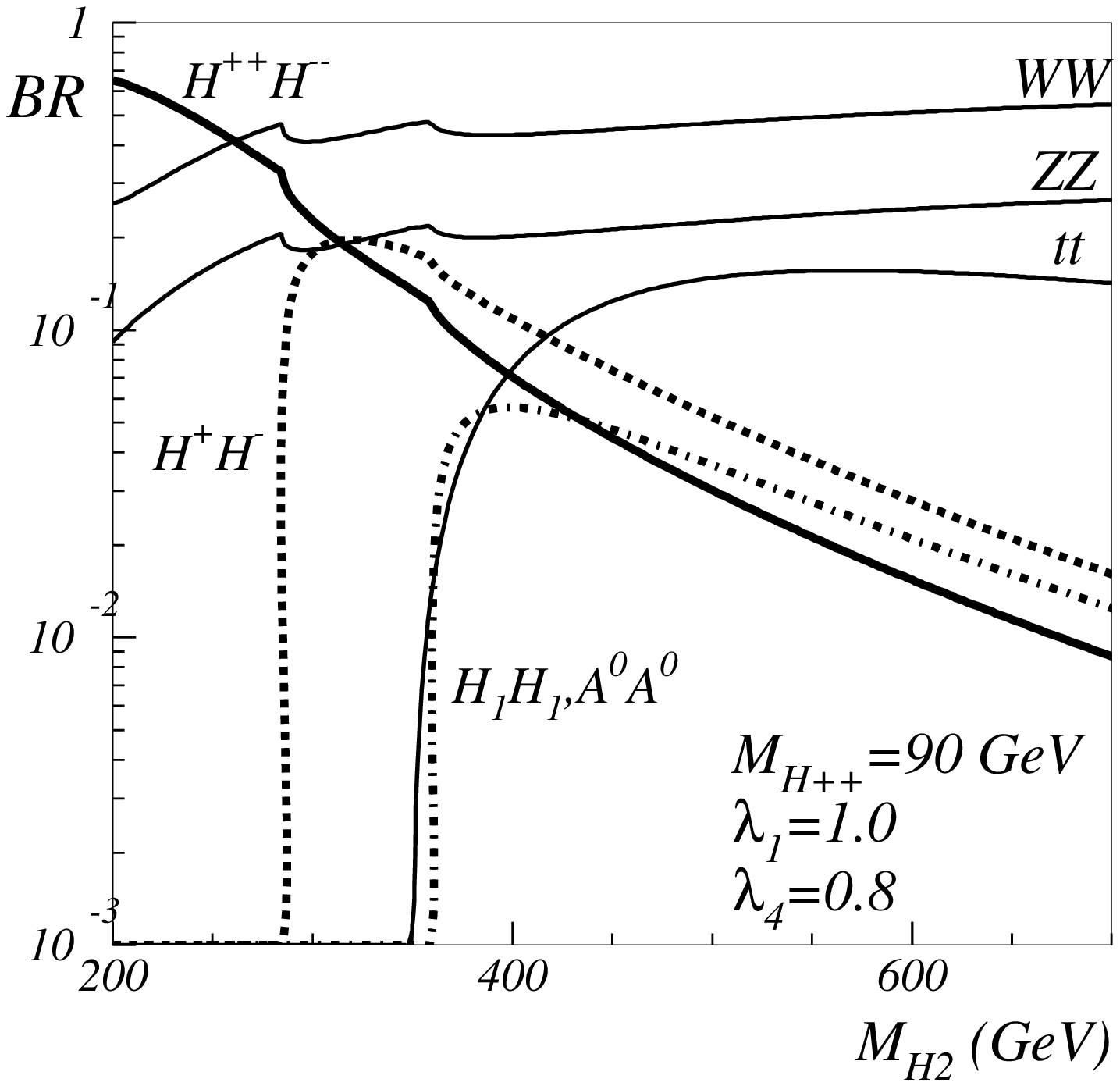}
\includegraphics[origin=c, angle=0, scale=0.46]{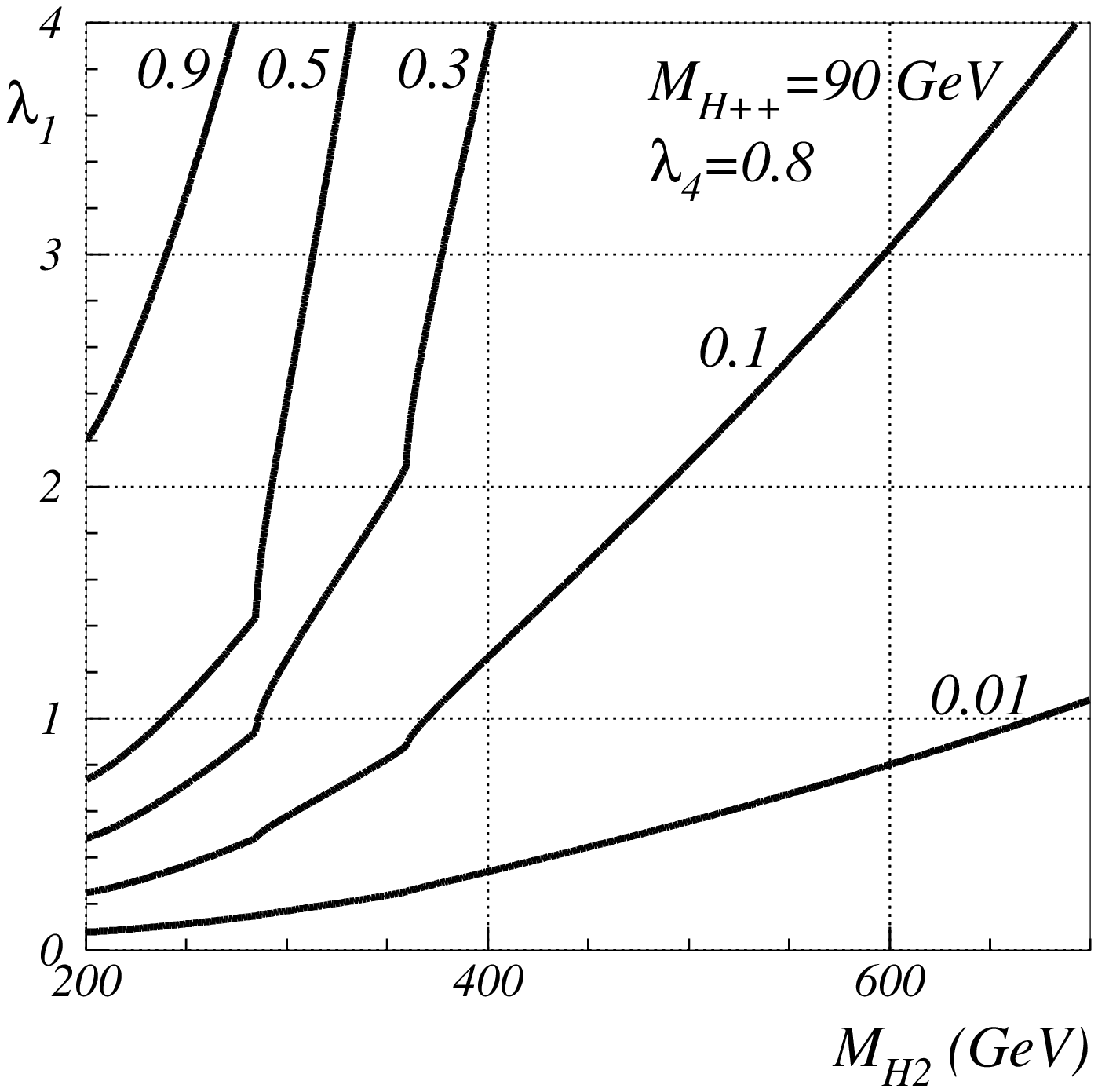}
\includegraphics[origin=c, angle=0, scale=0.46]{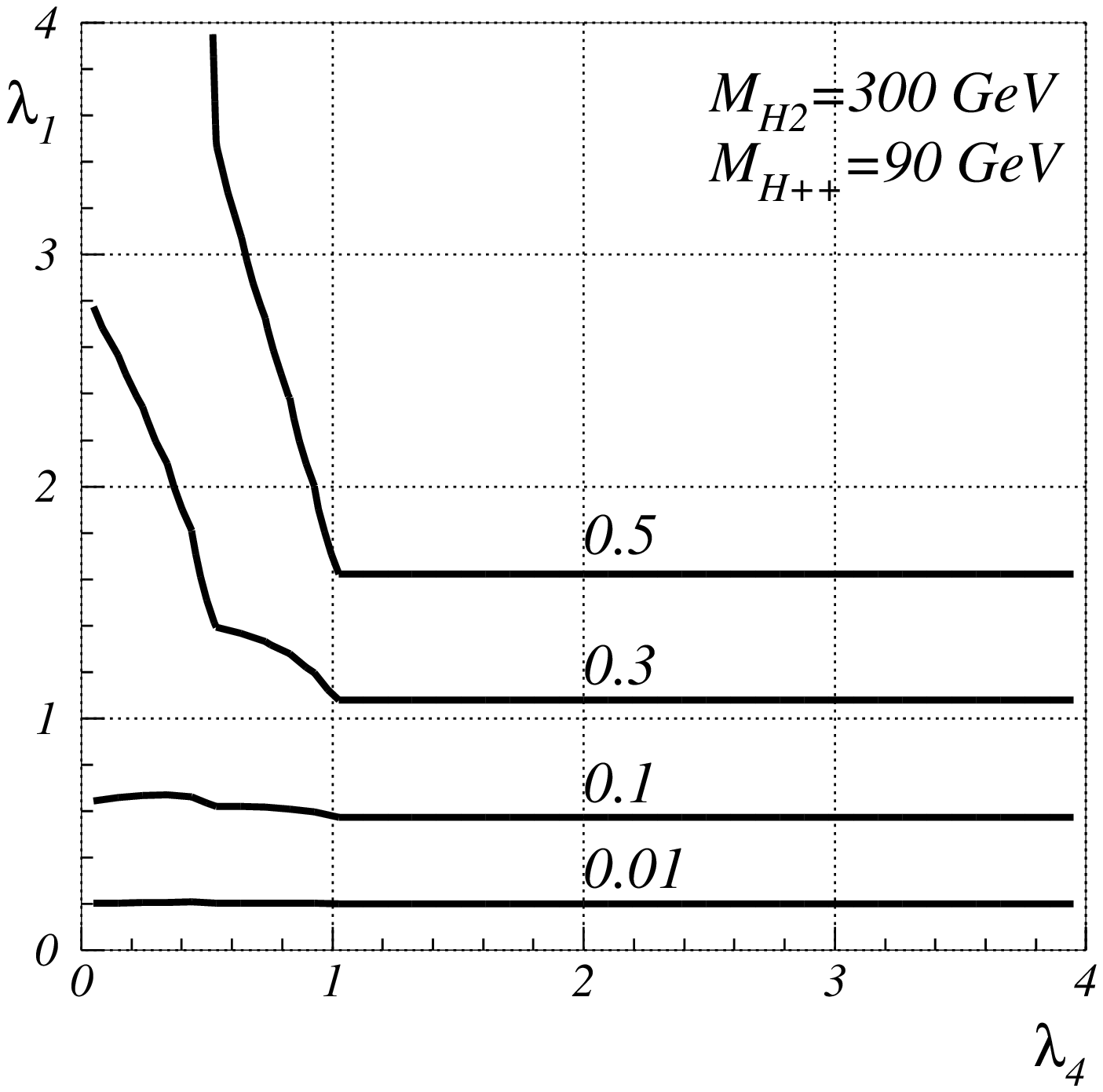}
\caption{Upper left panel a): branching ratios of $H_2$ as a function of $M_{H_2}$. Upper right panel b): Contours of 
BR$(H_2\to H^{++}H^{--}$) in the plane $[M_{H_2},\lambda_1]$. Lower panel c): Contours of BR$(H_2\to H^{++}H^{--}$
in the plane $[\lambda_4,\lambda_1]$. In all figures $M_{H^{\pm\pm}}=90$ GeV.
In a) and b) $\lambda_4=0.8$, which gives $M_{H^\pm}=142$ GeV and $M_{A^0,H_1}=179$ GeV.
In c) $M_{H_2}=300$ GeV.}
\label{fig.1}
\end{center}
\end{figure}

\begin{figure}[t]
\begin{center}
\includegraphics[origin=c, angle=0, scale=0.46]{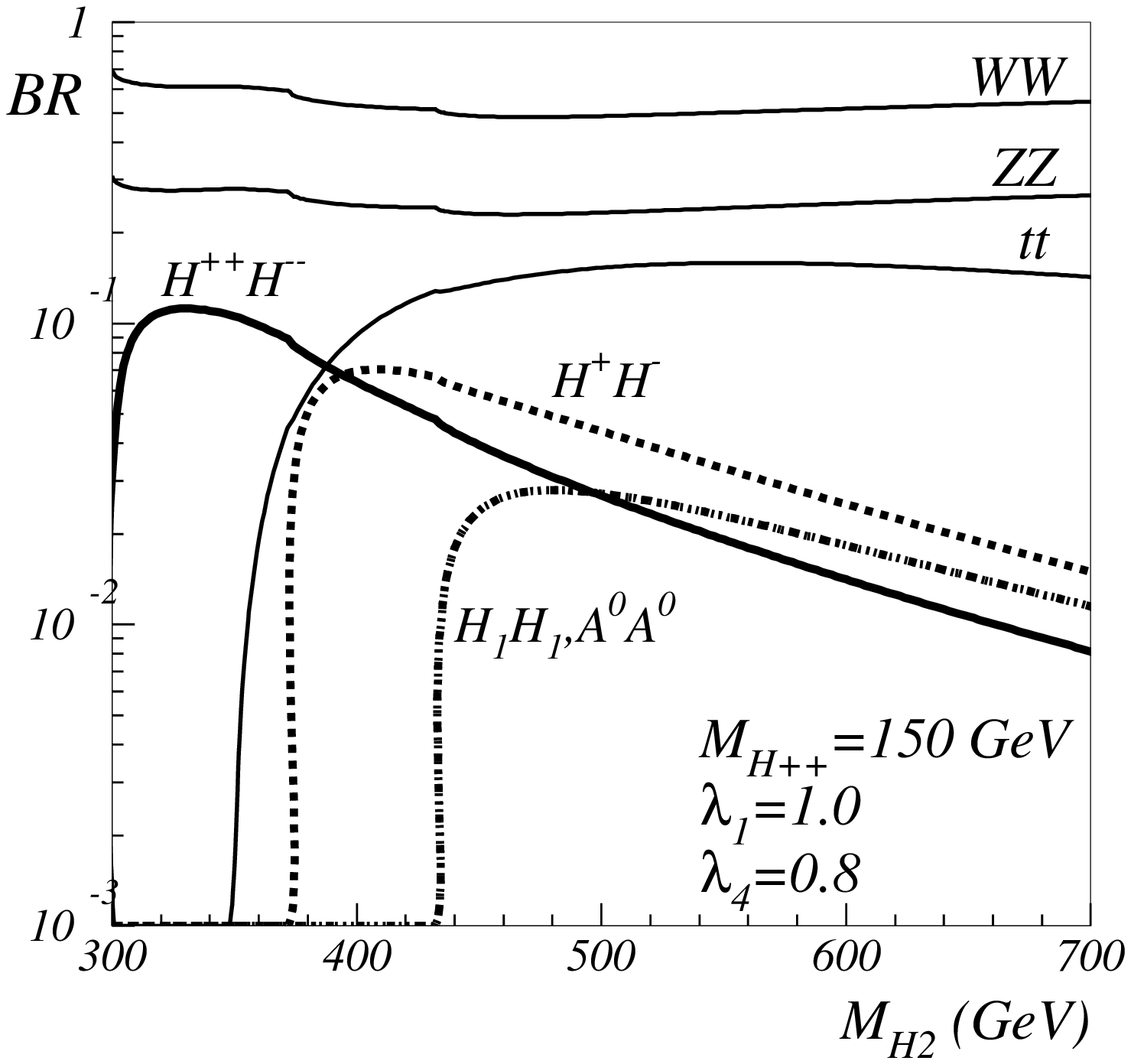}
\includegraphics[origin=c, angle=0, scale=0.46]{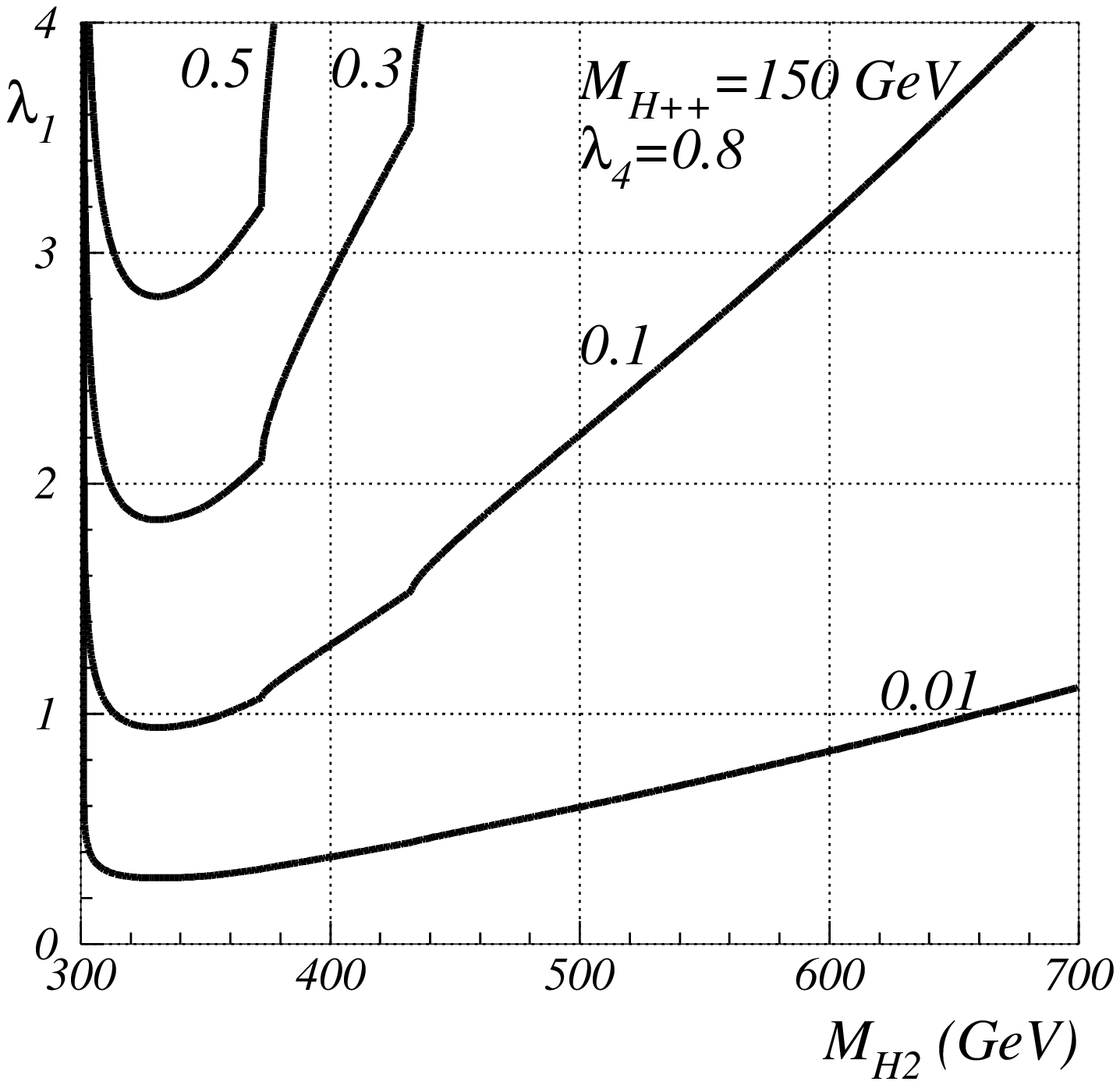}
\includegraphics[origin=c, angle=0, scale=0.46]{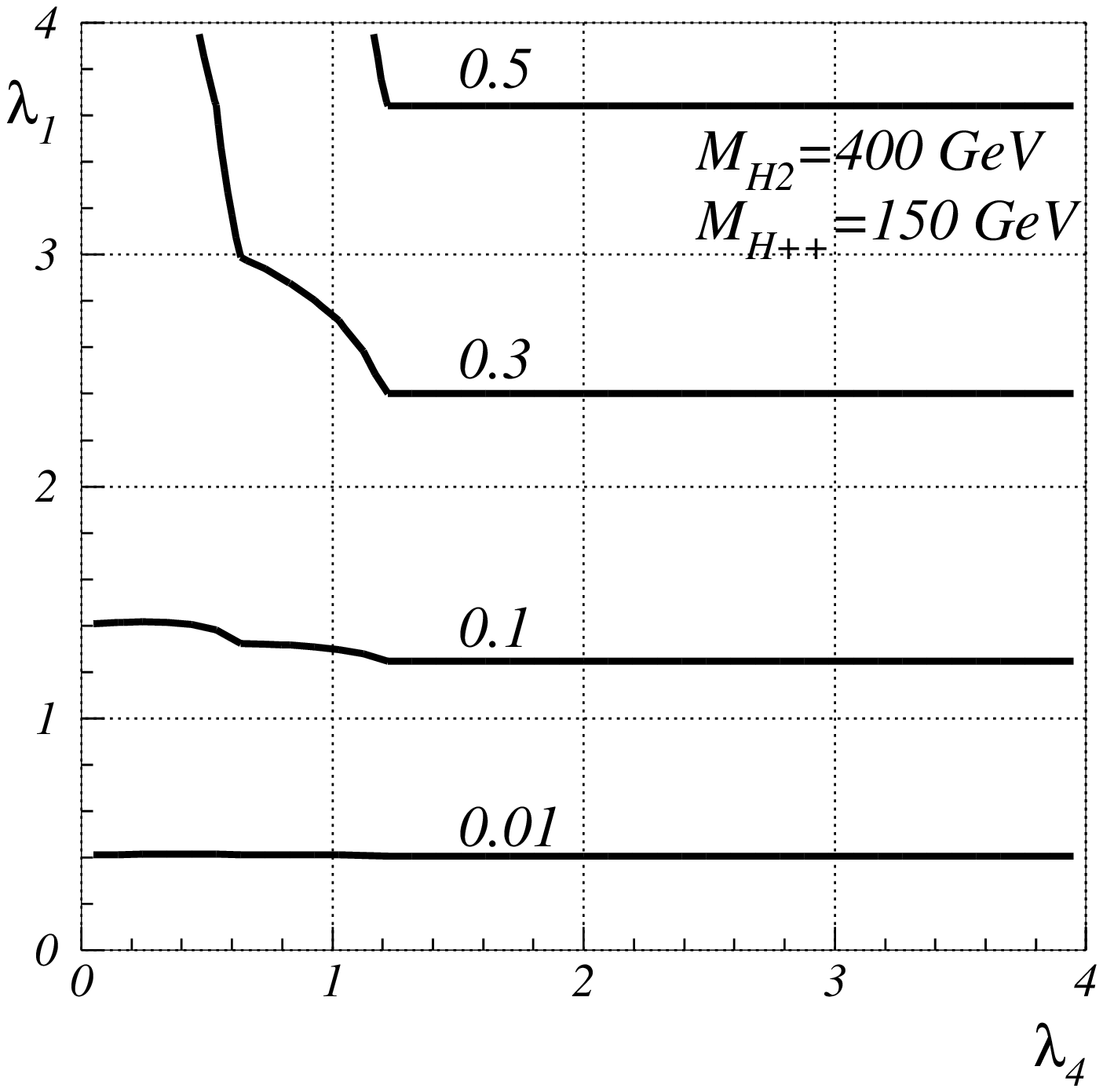}

\caption{Upper left panel a): branching ratios of $H_2$ as a function of $M_{H_2}$. Upper right panel b): Contours of 
BR$(H_2\to H^{++}H^{--}$) in the plane $[M_{H_2},\lambda_1]$. Lower panel c): Contours of BR$(H_2\to H^{++}H^{--}$
in the plane $[\lambda_4,\lambda_1]$. In all figures $M_{H^{\pm\pm}}=150$ GeV.
In a) and b) $\lambda_4=0.8$, which gives $M_{H^\pm}=186$ GeV and $M_{A^0,H_1}=216$ GeV.
In c) $M_{H_2}=400$ GeV.}
\label{fig.2}
\end{center}
\end{figure}

\begin{figure}[t]
\begin{center}
\includegraphics[origin=c, angle=0, scale=0.46]{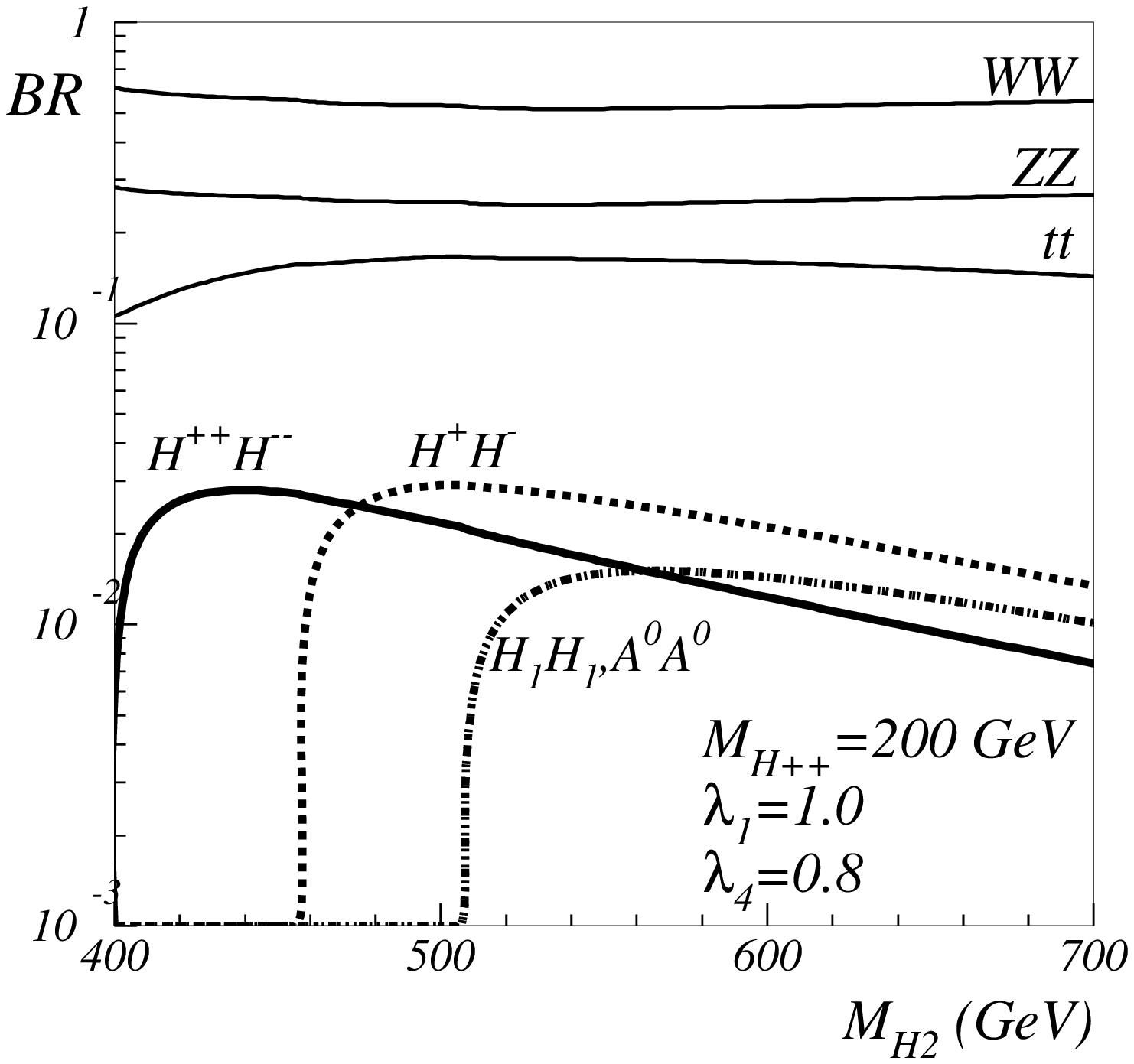}
\includegraphics[origin=c, angle=0, scale=0.46]{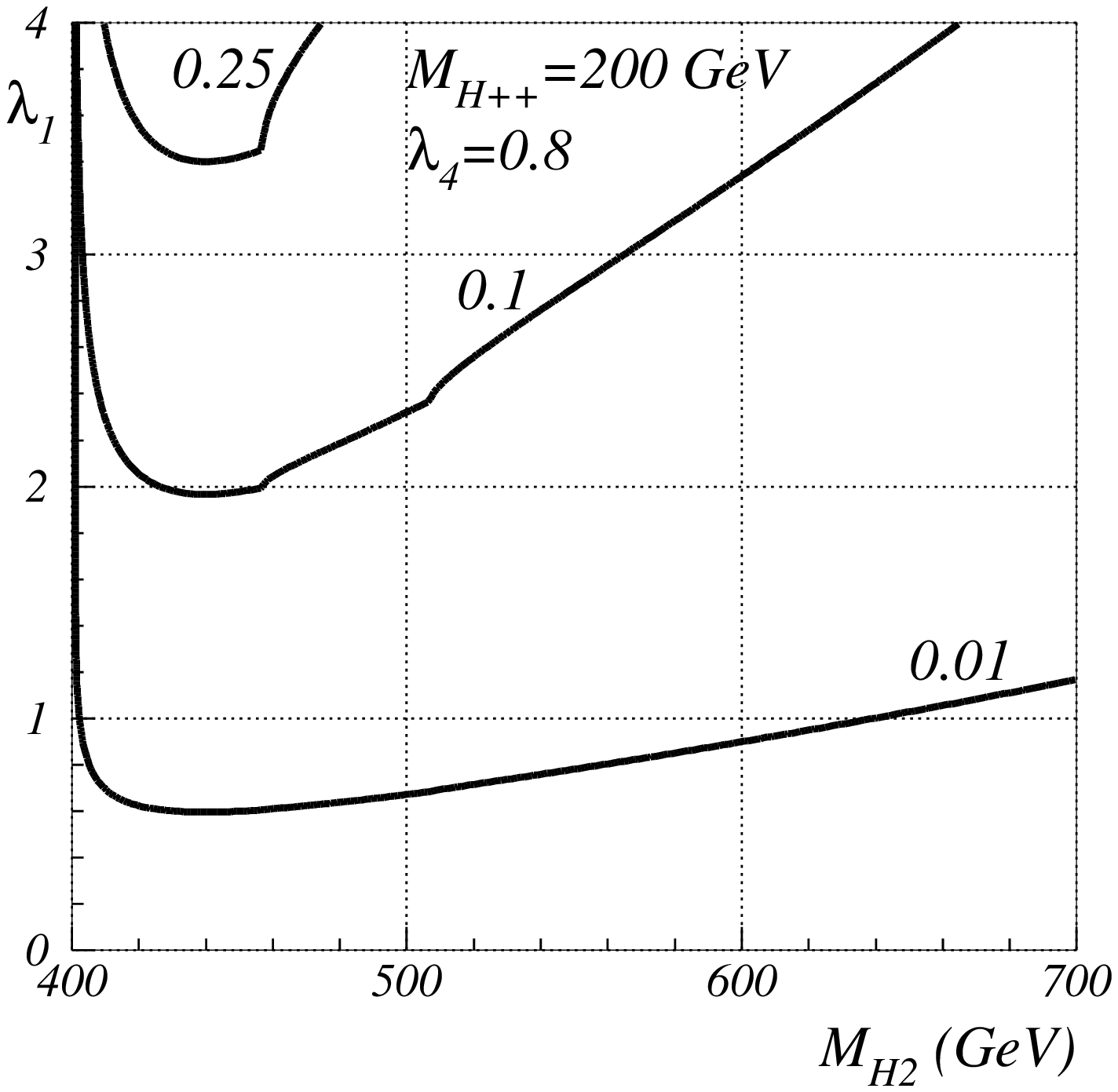}
\includegraphics[origin=c, angle=0, scale=0.46]{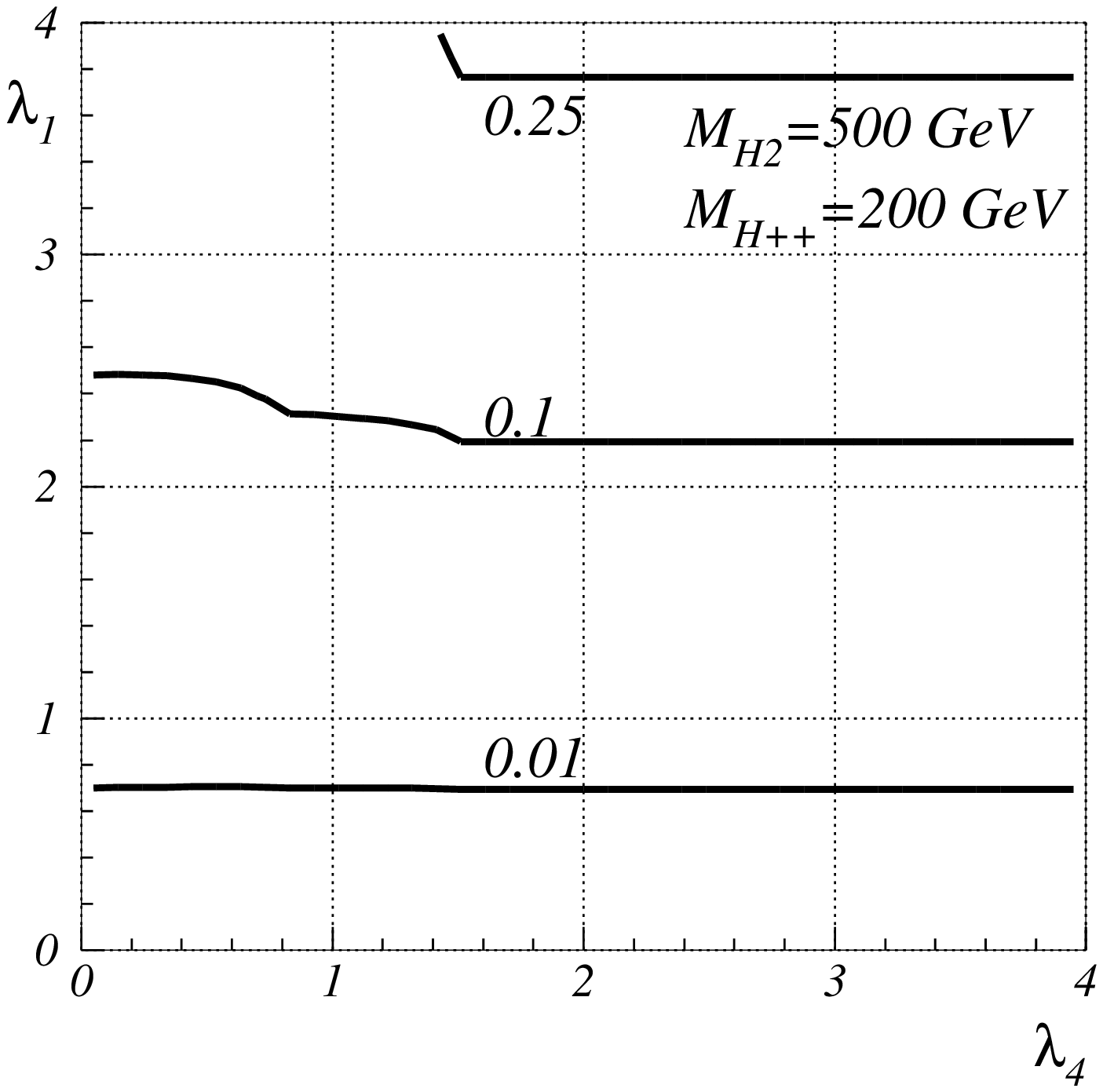}
\caption{Upper left panel a): branching ratios of $H_2$ as a function of $M_{H_2}$. Upper right panel b): Contours of 
BR$(H_2\to H^{++}H^{--}$) in the plane $[M_{H_2},\lambda_1]$. Lower panel c): Contours of BR$(H_2\to H^{++}H^{--}$
in the plane $[\lambda_4,\lambda_1]$. In all figures $M_{H^{\pm\pm}}=200$ GeV.
In a) and b) $\lambda_4=0.8$, which gives $M_{H^\pm}=228$ GeV and $M_{A^0,H_1}=253$ GeV.
In c) $M_{H_2}=500$ GeV.}
\label{fig.3}
\end{center}
\end{figure}

\begin{figure}[t]
\begin{center}
\includegraphics[origin=c, angle=0, scale=0.46]{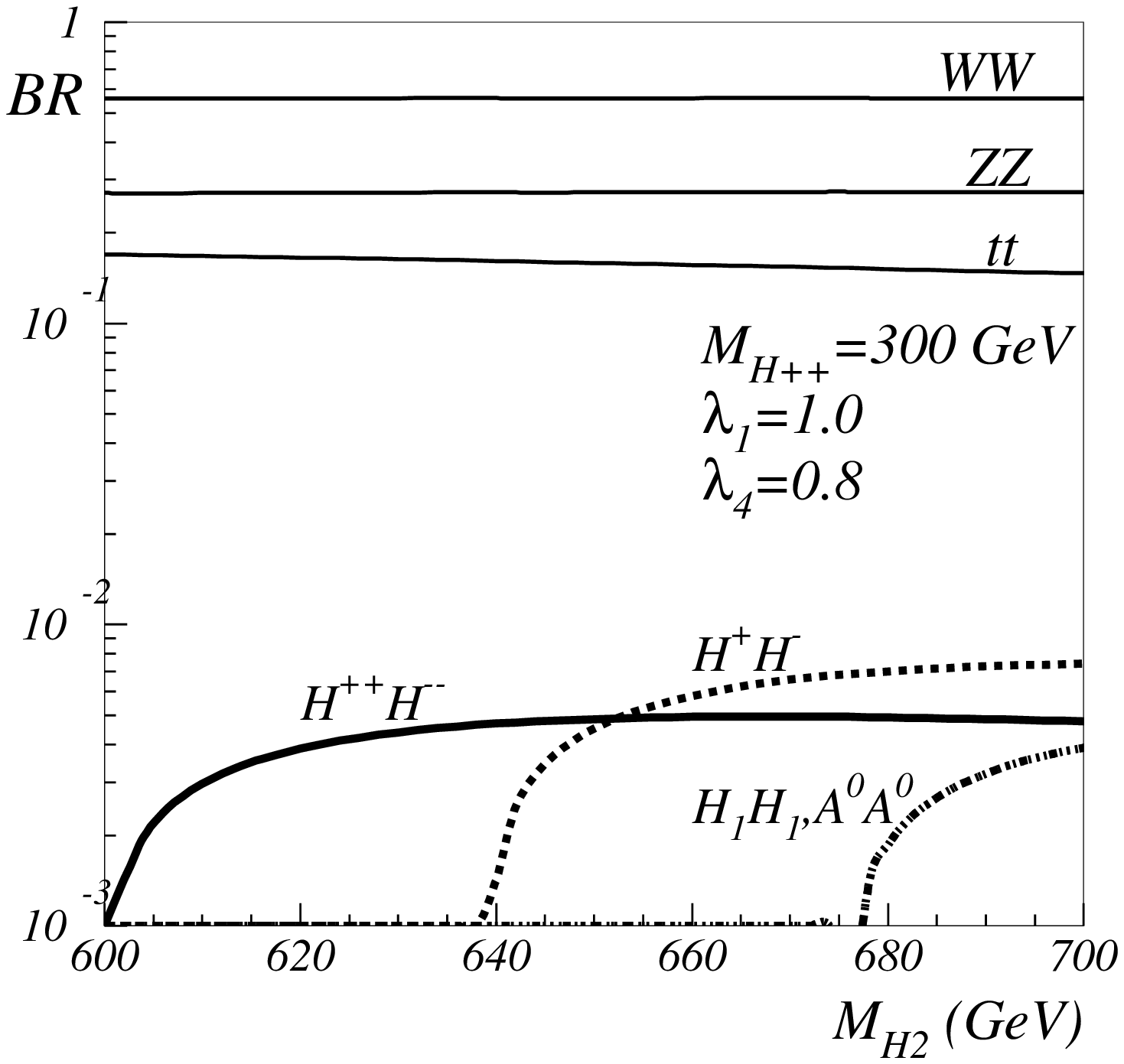}
\includegraphics[origin=c, angle=0, scale=0.46]{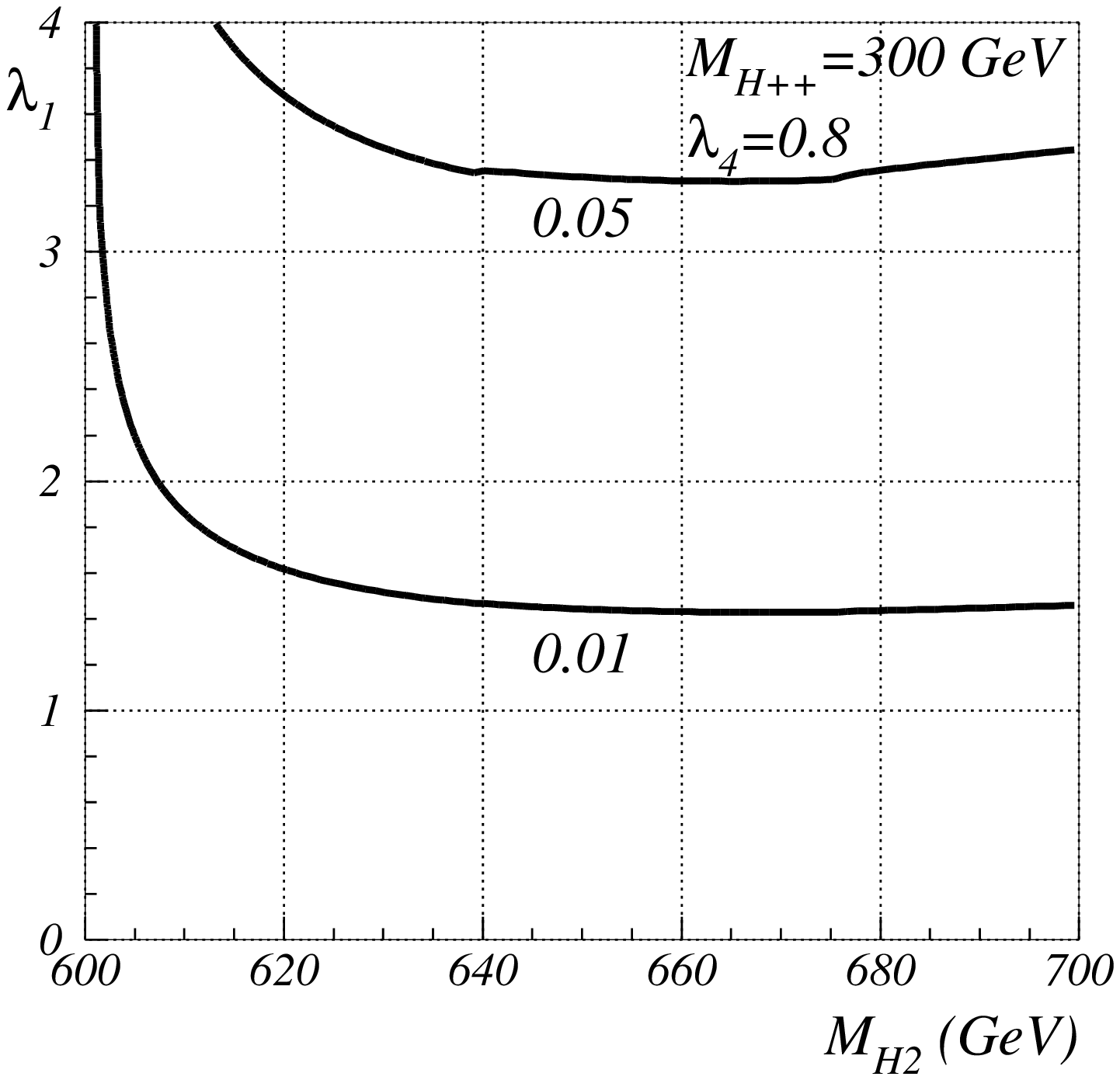}
\includegraphics[origin=c, angle=0, scale=0.46]{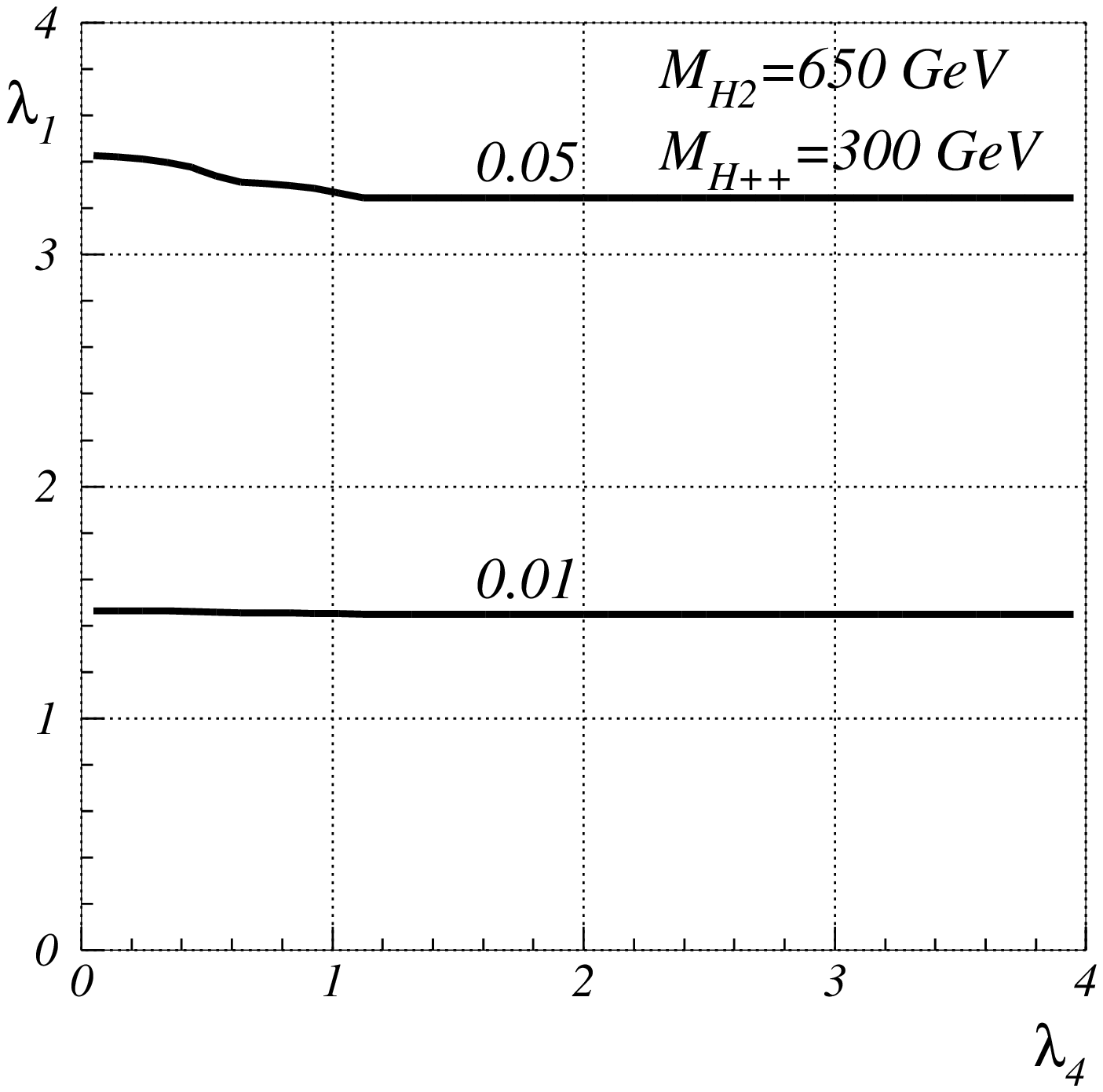}
\caption{Upper left panel a): branching ratios of $H_2$ as a function of $M_{H_2}$. Upper right panel b): Contours of 
BR$(H_2\to H^{++}H^{--}$) in the plane $[M_{H_2},\lambda_1]$. Lower panel c): Contours of BR$(H_2\to H^{++}H^{--}$
in the plane $[\lambda_4,\lambda_1]$. In all figures $M_{H^{\pm\pm}}=300$ GeV.
In a) and b) $\lambda_4=0.8$, which gives $M_{H^\pm}=320$ GeV and $M_{A^0,H_1}=338$ GeV.
In c) $M_{H_2}=650$ GeV.}
\label{fig.4}
\end{center}
\end{figure}

Fig.~(\ref{fig.1}a) shows the branching ratios of $H_2$ as a function of $M_{H_2}$.
We take $M_{H^{\pm\pm}}=90$ GeV, $\lambda=1$, and $\lambda_4=0.8$
(the latter choice gives $M_{H^\pm}=142$ GeV and $M_{A^0,H^0}=179$ GeV).
The magnitude of BR($H_2\to H^{++}H^{--}$) can reach $65\%$ for
$M_{H_2}=200$ GeV, and stays as the dominant channel until
$M_{H_2}\sim 260$ GeV, at which $H_2\to WW$ becomes dominant.
BR($H_2\to H^{++}H^{--}$) falls below
BR($H_2\to ZZ$) at around $M_{H_2}=320$ GeV. This dependence on $M_{H_2}$ can be explained by the fact 
that the partial widths of $H_2\to WW,ZZ$ are proportional to
$M^3_{H_2}$, and so ultimately these channels will dominate for
larger $M_{H_2}$. The other decays of $H_2$ to two triplet scalars also can have sizeable branching ratios,  
with BR($H_2\to H^{+}H^{-}$) reaching $20\%$ at most, and exceeds BR($H_2\to H^{++}H^{--}$) for $M_{H_2}\gsim 315$ GeV. 
The branching ratios of $H_2\to A^0A^0$ and $H_2\to H_1H_1$ are equal; they are plotted 
individually and their sum  
peaks at $\sim 10\%$.
In fig.~(\ref{fig.1}b) we show contours of 
BR$(H_2\to H^{++}H^{--}$) in the plane $[M_{H_2},\lambda_1]$. 
As expected, BR$(H_2\to H^{++}H^{--}$) takes its largest values for 
large $\lambda_1$ and light $M_{H_2}$, with BR$(H_2\to H^{++}H^{--})> 90\%$ being
possible. We note that such a scenario would render the 
searches for $H_2\to WW,ZZ$ ineffective until a very large amount of
integrated luminosity is obtained.
In fig.~(\ref{fig.1}c) we show contours of BR$(H_2\to H^{++}H^{--}$)
in the plane $[\lambda_4,\lambda_1]$, fixing $M_{H_2}=300$ GeV.
For $\lambda_4\gsim 1$ only the decay $H_2\to H^{++}H^{--}$ is open kinematically and
so the contours are horizontal. For $\lambda_4=0$ all the triplet scalars are degenerate
and thus all four decay channels are open. Figs.~\ref{fig.2},\ref{fig.3} and \ref{fig.4} are analogies of fig.~\ref{fig.1}, but with 
$M_{H^{\pm\pm}}=150$ GeV, 200 GeV and 300 GeV respectively. In all figures we take
$M_{H_2} > 2M_{H^{\pm\pm}}$. One can see similar qualitative behaviour, but since 
the lowest value of $M_{H_2}$ is larger 
in figs.~\ref{fig.2},\ref{fig.3} and \ref{fig.4} than in fig.~\ref{fig.1}, the maximum values of BR$(H_2\to H^{++}H^{--})$ are less than in fig.1. 
However, in fig.~2b, fig.~3b and fig.~4b it can be seen 
that BR$(H_2\to H^{++}H^{--})>50\%,25\%,5\%$ respectively is possible for $\lambda_1\gsim 3$. 

It clear that BR$(H_2\to H^{++}H^{--})$ can be sizeable, and we will now quantify the 
magnitude of the pair production of $H^{\pm\pm}$ which originates from production and decay of $H_2$.
At hadron colliders $H_2$ is dominantly created via gluon-gluon fusion, $gg \to H_2$. 
For $M_{H_2}=2M_{H^{\pm\pm}}$ the cross section of
$gg \to H_2$ at the LHC is significantly larger than the cross section for
the direct production mechanisms of $H^{\pm\pm}$ (i.e. $q\overline q \to \gamma^*,Z^*\to H^{++}H^{--}$
and  $q'\overline q \to W \to H^{\pm\pm}H^{\mp}$). However, the same is not true at the Tevatron,
and $\sigma(gg \to H_2)\lsim \sigma(q\overline q\to \gamma^*,Z^*\to H^{++}H^{--})$ for 
$M_{H_2}>2M_{H^{\pm\pm}}$.

\begin{figure}[t]
\begin{center}
\includegraphics[origin=c, angle=0, scale=0.46]{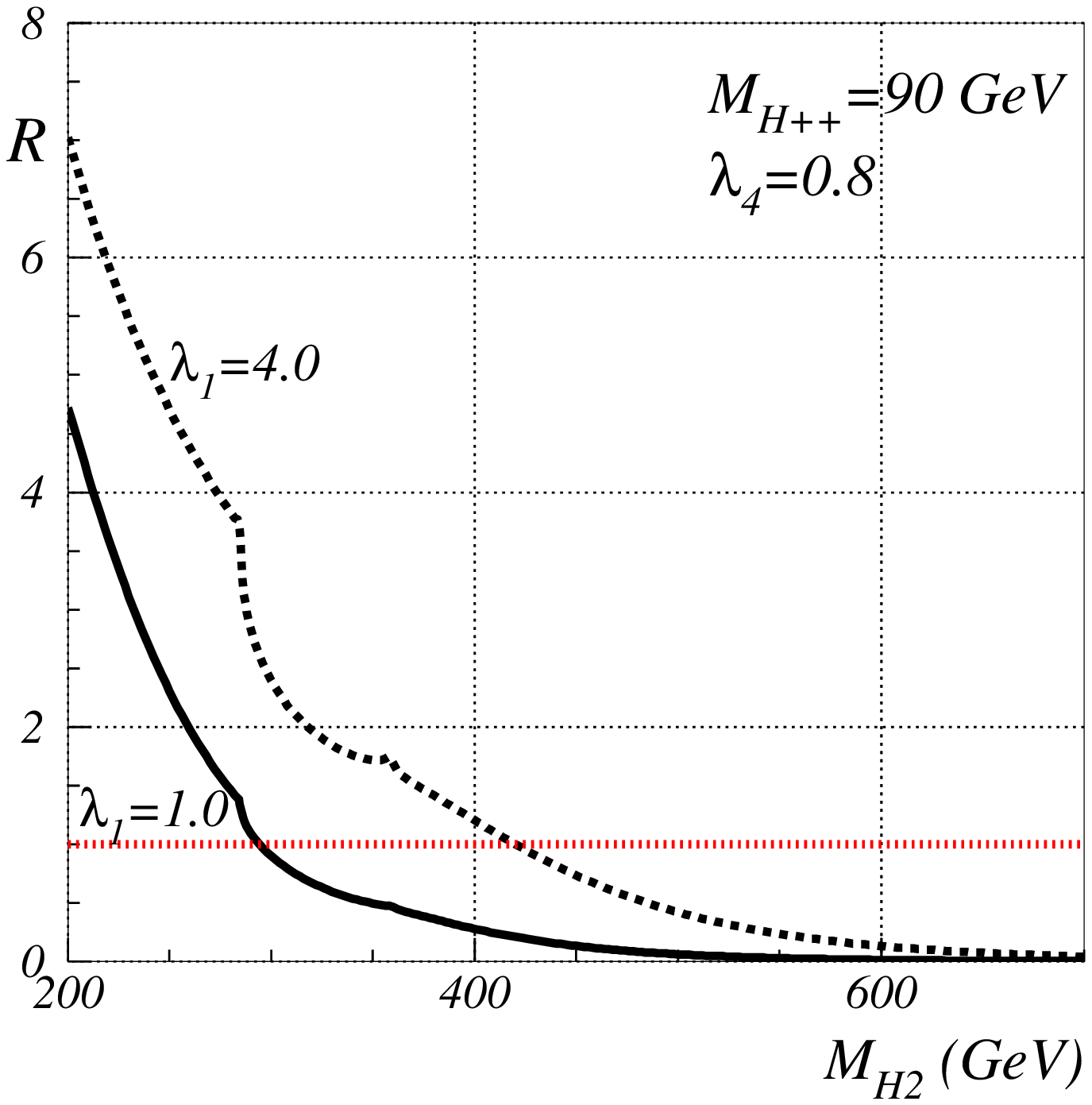}
\includegraphics[origin=c, angle=0, scale=0.46]{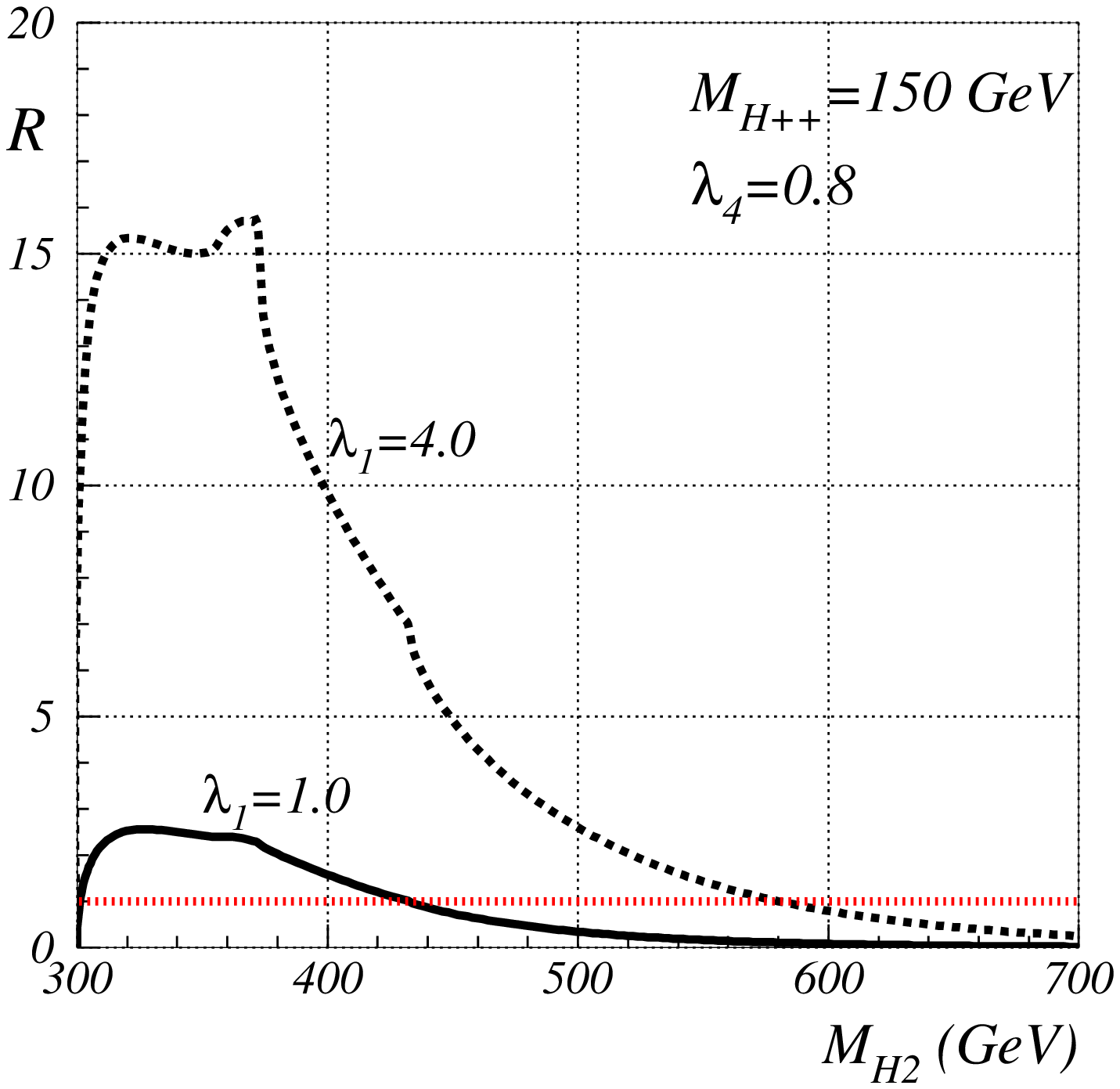}
\includegraphics[origin=c, angle=0, scale=0.46]{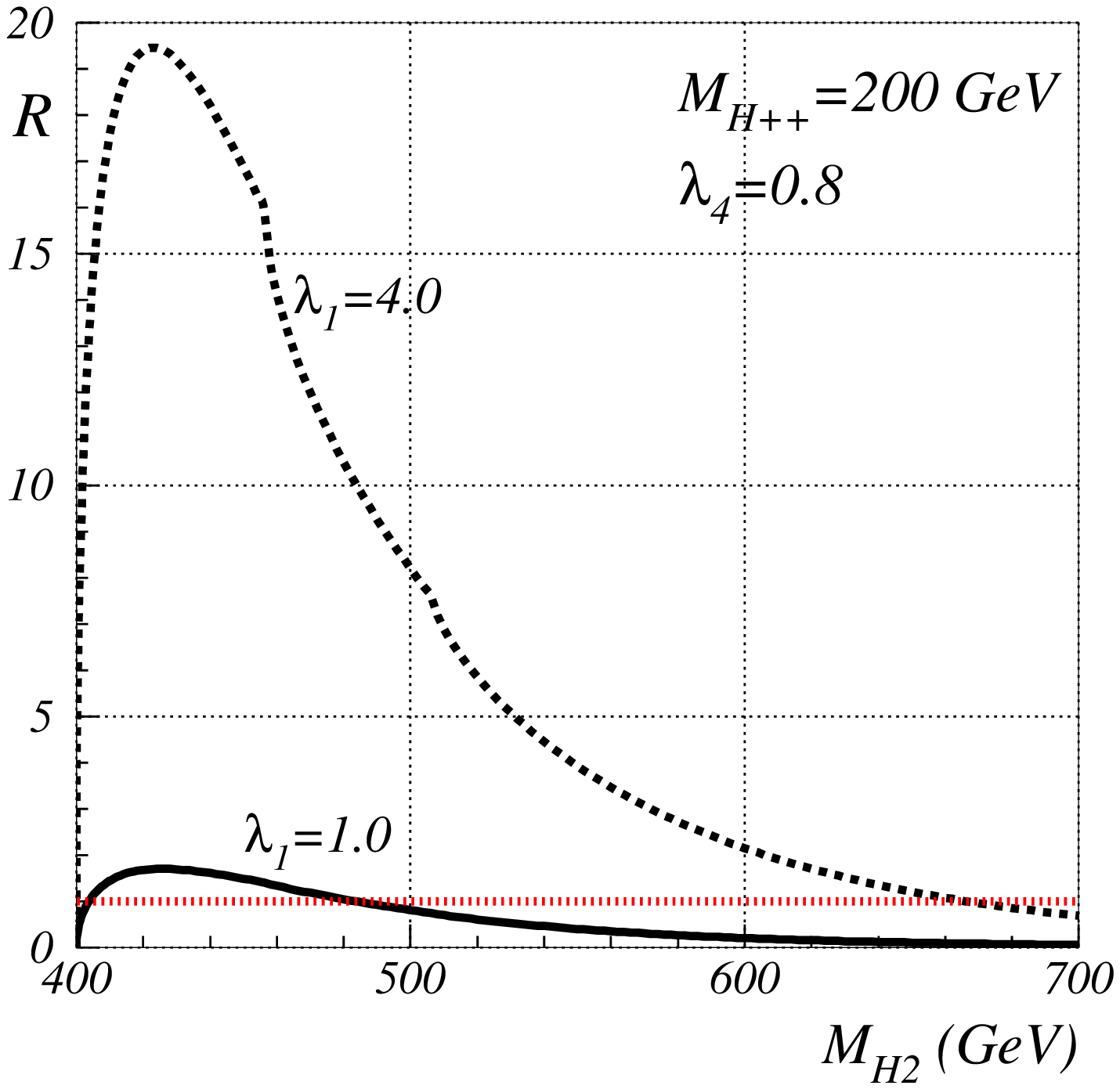}
\includegraphics[origin=c, angle=0, scale=0.46]{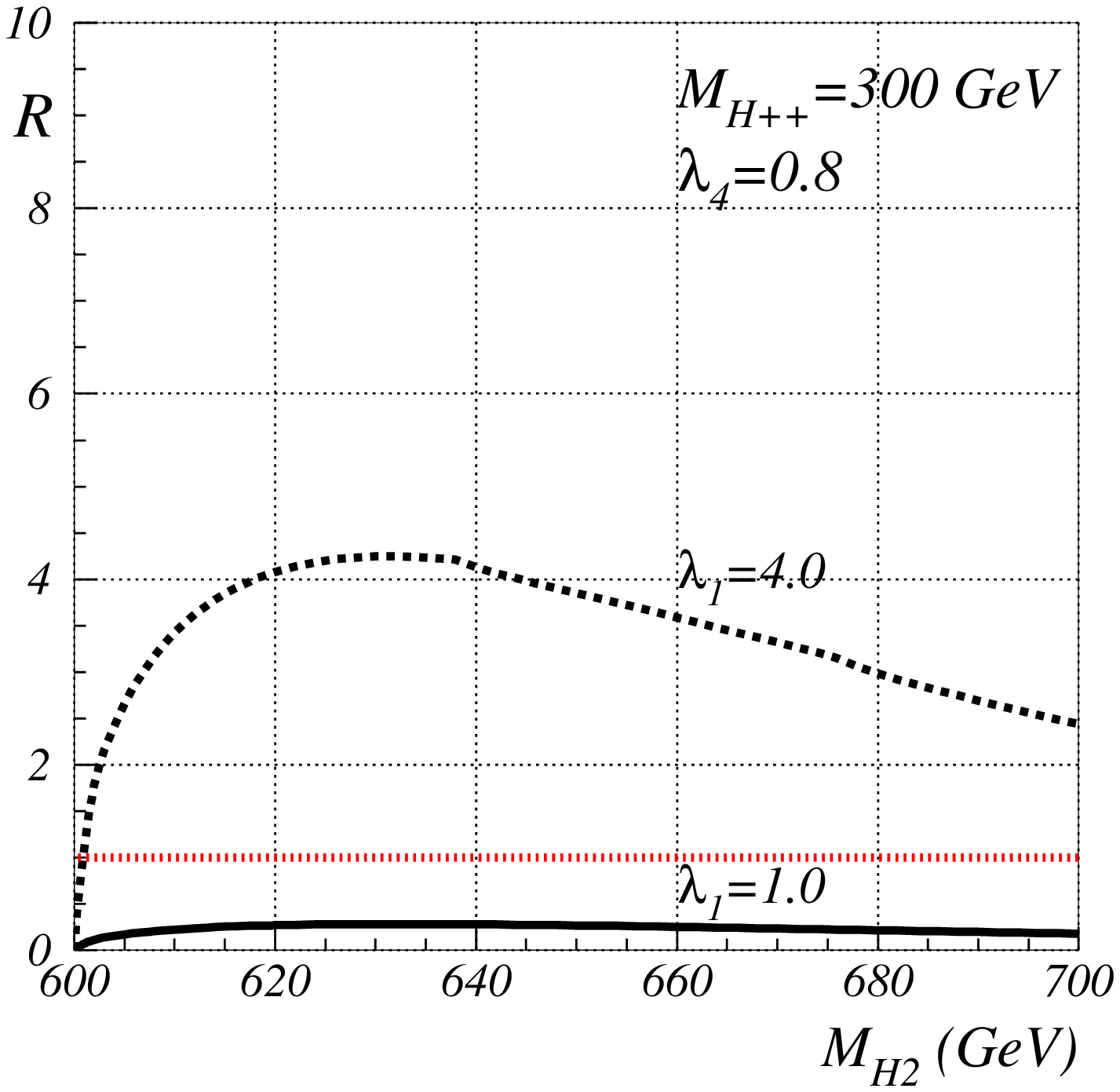}
\caption{The ratio $R=\sigma(gg \to H_2)\times {\rm BR} (H_2\to H^{++}H^{--})/
\sigma(q\overline q\to H^{++}H^{--})$
at the LHC (with $\sqrt s=14$ TeV) as a function of $M_{H_2}$.
The two curves are for $\lambda_1=1$ and $\lambda_1=4$, with $\lambda_4=0.8$ .
We take $M_{H^{\pm\pm}}=90$ GeV in panel (a), $M_{H^{\pm\pm}}=150$ GeV in panel (b),
$M_{H^{\pm\pm}}=200$ GeV in panel (c), and $M_{H^{\pm\pm}}=300$ GeV in panel (d).
The horizontal line shows $R=1$.}
\label{fig.5}
\end{center}
\end{figure}

We introduce the ratio $R$, defined by:
\begin{eqnarray}
R=\frac{\sigma(gg \to H_2)\times {\rm BR} (H_2\to H^{++}H^{--})}{\sigma(q\overline q\to \gamma^*,Z^*\to H^{++}H^{--})}
\label{ratio-R}
\end{eqnarray}
The denominator in eq.~(\ref{ratio-R}) is the conventional mechanism for production of  $H^{++}H^{--}$,
which is assumed in the ongoing searches for $H^{\pm\pm}$. The numerator is a novel mechanism which
contributes when ${\rm BR} (H_2\to H^{++}H^{--})\ne 0$.
We will now study the magnitude of the ratio $R$ at the LHC (with $\sqrt s=7$ TeV and 14 TeV) and at the Tevatron.
In Fig.~\ref{fig.5} we plot $R$ as a function of $M_{H_2}$ at the LHC with $\sqrt s=14$ TeV, for $M_{H^{\pm\pm}}=90$ GeV,
150 GeV, 200 GeV and 300 GeV.
The factorisation scale and normalisation scale are both taken to be $M_{H_2}$ for $gg \to H_2$, while for
$q\overline q\to \gamma^*,Z^*\to H^{++}H^{--}$ both scales are taken to be the partonic centre-of-mass energy. 
We use CTEQ6L1 parton distribution functions \cite{Pumplin:2002vw} with the leading-order 
partonic cross section for $gg\to H_2$ \cite{Georgi:1977gs}.
We do not apply 
QCD $K$ factors which, would increase the value of $R$ because the ratio of the $K$ factors for
$\sigma(gg \to H_2)$ \cite{Djouadi:2005gi} and $\sigma(q\overline q\to \gamma^*,Z^*\to H^{++}H^{--})$ \cite{Muhlleitner:2003me} 
is about 1.4 in the region of
interest of $M_{H_2}$ and  $M_{H^{\pm\pm}}$.

In fig.~(\ref{fig.5}a) we take $M_{H^{\pm\pm}}=90$ GeV, which fixes the value
of $\sigma(q\overline q\to H^{++}H^{--}$), and $\lambda_4=0.8$.
We take $\lambda_1=1$ and 4.
If $\lambda_1=1$ one can see that $R=4.7$ for $M_{H_2}=200$ GeV, and 
$R>1$ for $M_{H_2}<290$ GeV.
If $\lambda_1=4$, one finds that $R=7.0$ for $M_{H_2}=200$ GeV, and 
$R>1$ for $M_{H_2}<420$ GeV. The noticeable drop in the value of $R$ for $M_{H_2}\sim 280$ GeV
is due to the opening of the decay channel $H_2\to H^{+}H^{-}$ (see fig.~(\ref{fig.1}a)).
Both $\sigma(gg \to H_2)$ and ${\rm BR} (H_2\to H^{++}H^{--})$ are decreasing functions 
of $M_{H_2}$, which explains the overall dependence of $R$ on $M_{H_2}$.
Note that $R$ does not fall so sharply with $M_{H_2}$ in the region $320\,\, {\rm GeV} < M_{H_2} < 380$ GeV, because
$\sigma(gg \to H_2)$ increases in magnitude up to a local maximum at $M_{H_2}=2m_t$, before decreasing again.
In fig.~(\ref{fig.5}b) we take $M_{H^{\pm\pm}}=150$ GeV, and $R\sim 16$ for $M_{H_2}=2m_t$ and $\lambda_1=4$. 
Larger values of $R$ are attainable because the magnitude of $\sigma(q\overline q\to \gamma^*,Z^*\to H^{++}H^{--})$
(i.e. the denominator eq.~(\ref{ratio-R})) diminishes considerably when going from $M_{H^{\pm\pm}}=90$ GeV to $M_{H^{\pm\pm}}=150$ GeV, 
while the corresponding decrease in $\sigma(gg \to H_2)$ for larger $M_{H_2}$ is relatively less. In
fig.~(\ref{fig.5}c) (for $M_{H^{\pm\pm}}=200$ GeV) the maximum value is $R\sim 19$, and in 
fig.~(\ref{fig.5}d) the maximum value is $R\sim 4$. It is evident that there is a sizeable parameter space for
$R>1$, and thus $gg \to H_2$ could give a significant (or even dominant) contribution to the 
pair production of $H^{\pm\pm}$ at the LHC. We also note that the decay $H_2\to H^{+}H^{-}$ 
(which can have a large BR in fig.~(\ref{fig.1}a) $\to$ fig.~(\ref{fig.4}a)) can lead to additional production 
of $H^{++}H^{--}$ because the branching ratio of the decay $H^\pm \to H^{\pm\pm}W^*$ can be large in a sizeable
parameter space of $[v_\Delta,M_{H^{\pm}}-M_{H^{\pm\pm}}$], as shown in \cite{Akeroyd:2011zz}. 
In fact, in fig~.(\ref{fig.1}a) $\to$ fig.~(\ref{fig.4}a) the mass splitting ($M_{H^{\pm}}-M_{H^{\pm\pm}}$) is between
20 GeV and 52 GeV, and with our chosen value of $v_\Delta=10^{-2}$ MeV one would have
BR($H^\pm \to H^{\pm\pm}W^*)>99\%$.

\begin{figure}[t]
\begin{center}
\includegraphics[origin=c, angle=0, scale=0.46]{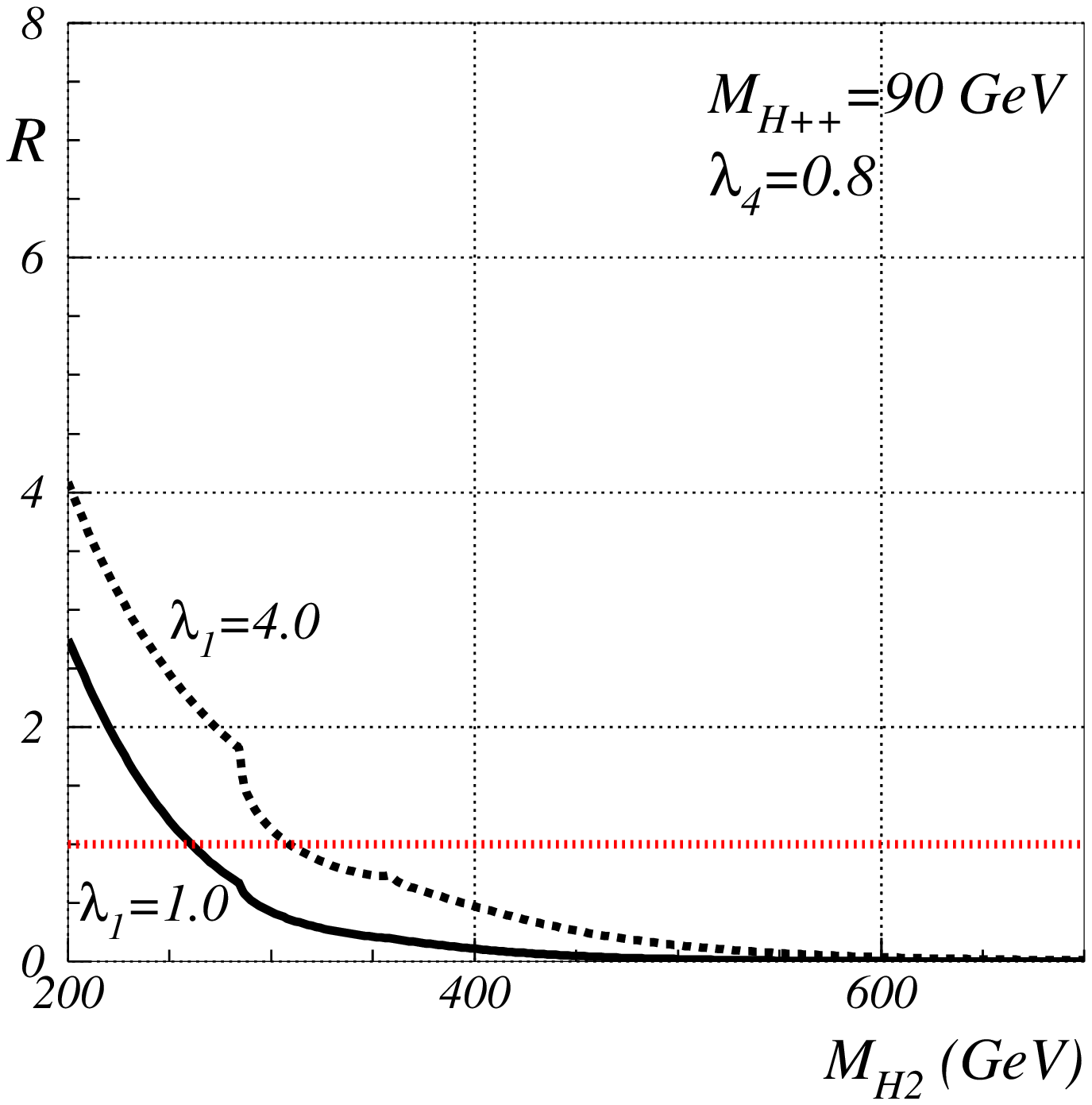}
\includegraphics[origin=c, angle=0, scale=0.46]{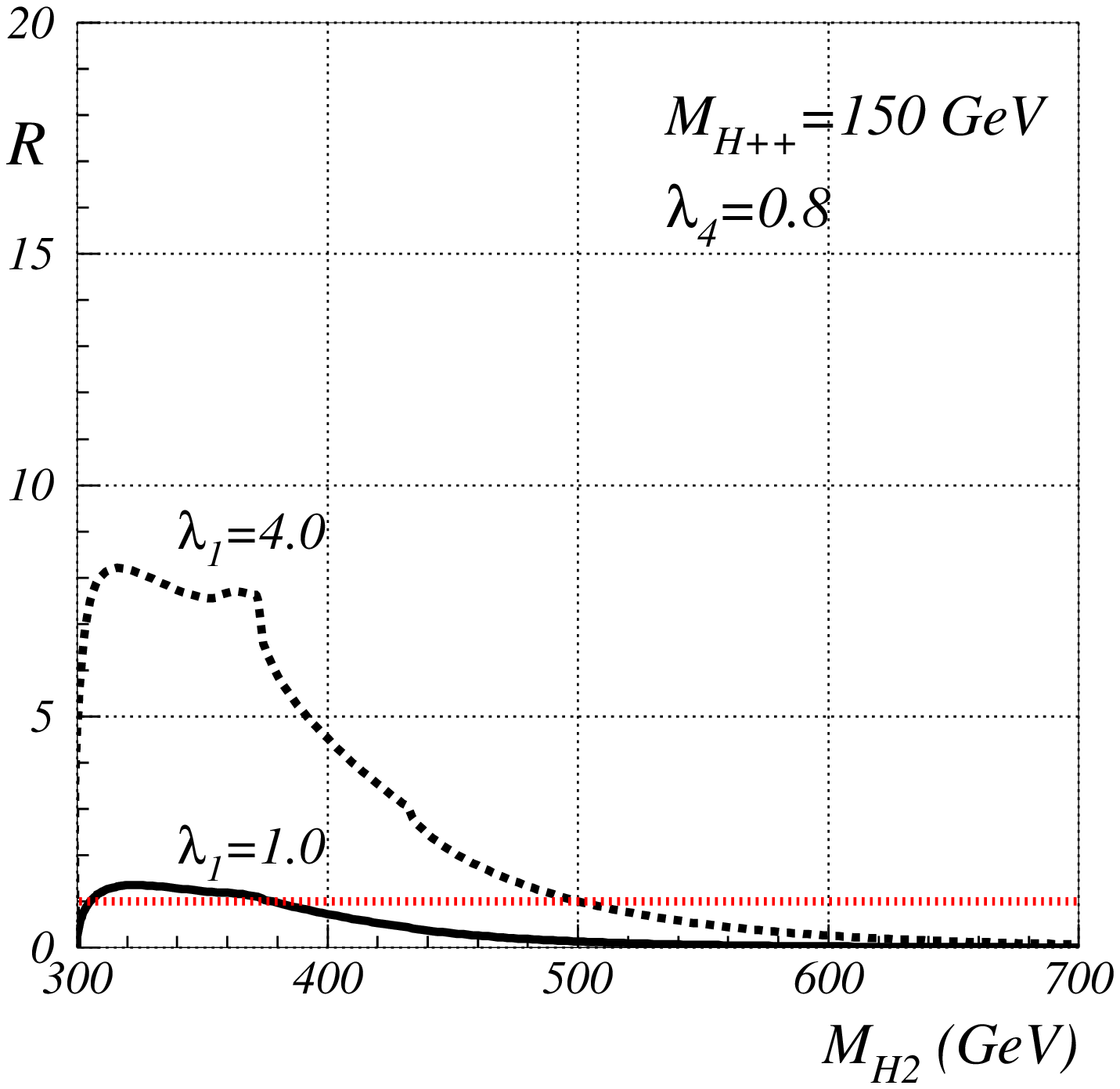}
\includegraphics[origin=c, angle=0, scale=0.46]{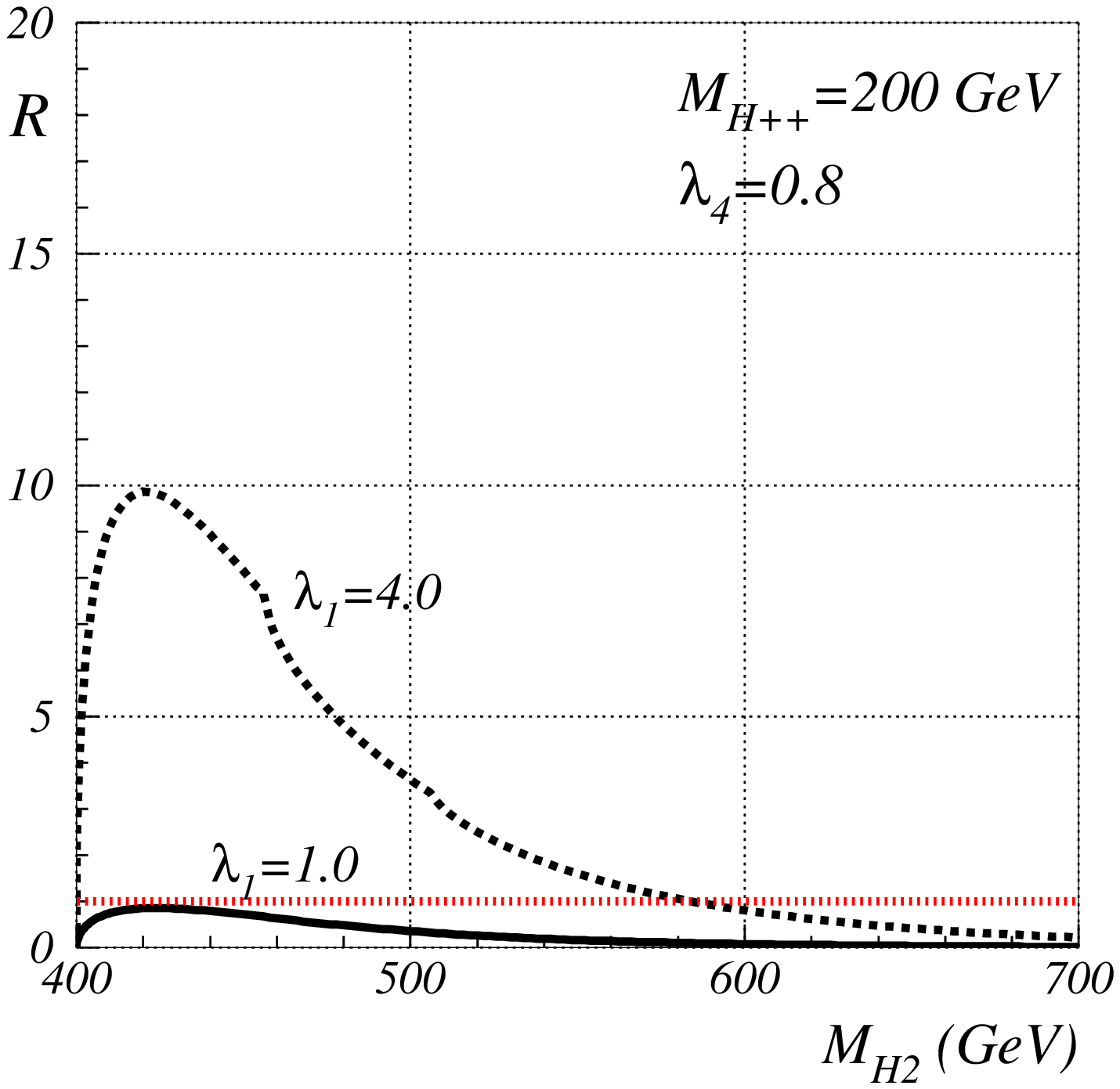}
\includegraphics[origin=c, angle=0, scale=0.46]{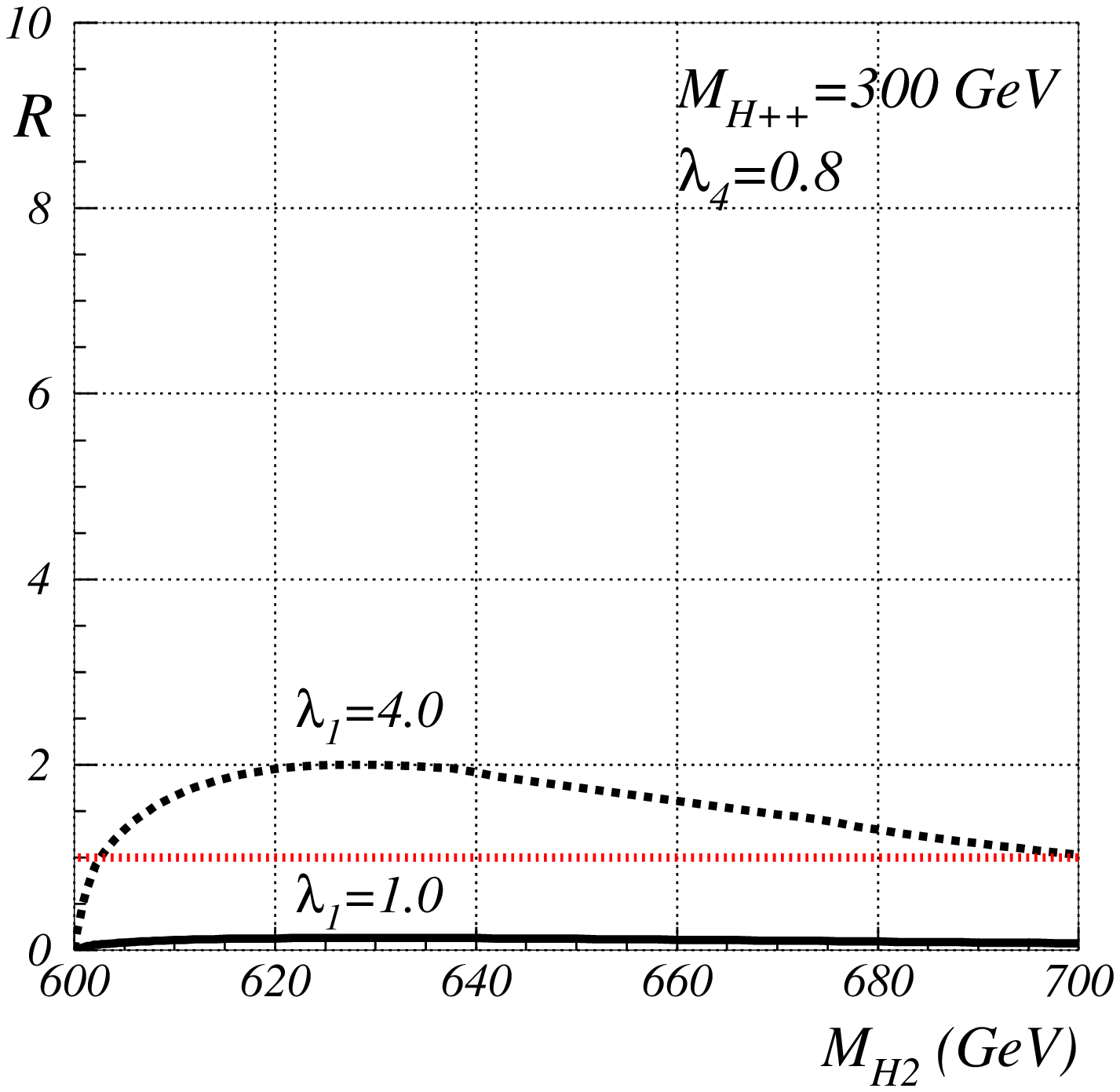}
\caption{The ratio $R=\sigma(gg \to H_2)\times {\rm BR} (H_2\to H^{++}H^{--})/
\sigma(q\overline q\to H^{++}H^{--})$
at the LHC (with $\sqrt s=7$ TeV) as a function of $M_{H_2}$.
The two curves are for $\lambda_1=1$ and $\lambda_1=4$, with $\lambda_4=0.8$ .
We take $M_{H^{\pm\pm}}=90$ GeV in panel (a), $M_{H^{\pm\pm}}=150$ GeV in panel (b),
$M_{H^{\pm\pm}}=200$ GeV in panel (c), and $M_{H^{\pm\pm}}=300$ GeV in panel (d).
The horizontal line shows $R=1$.}
\label{fig.6}
\end{center}
\end{figure}

\begin{figure}[t]
\begin{center}
\includegraphics[origin=c, angle=0, scale=0.46]{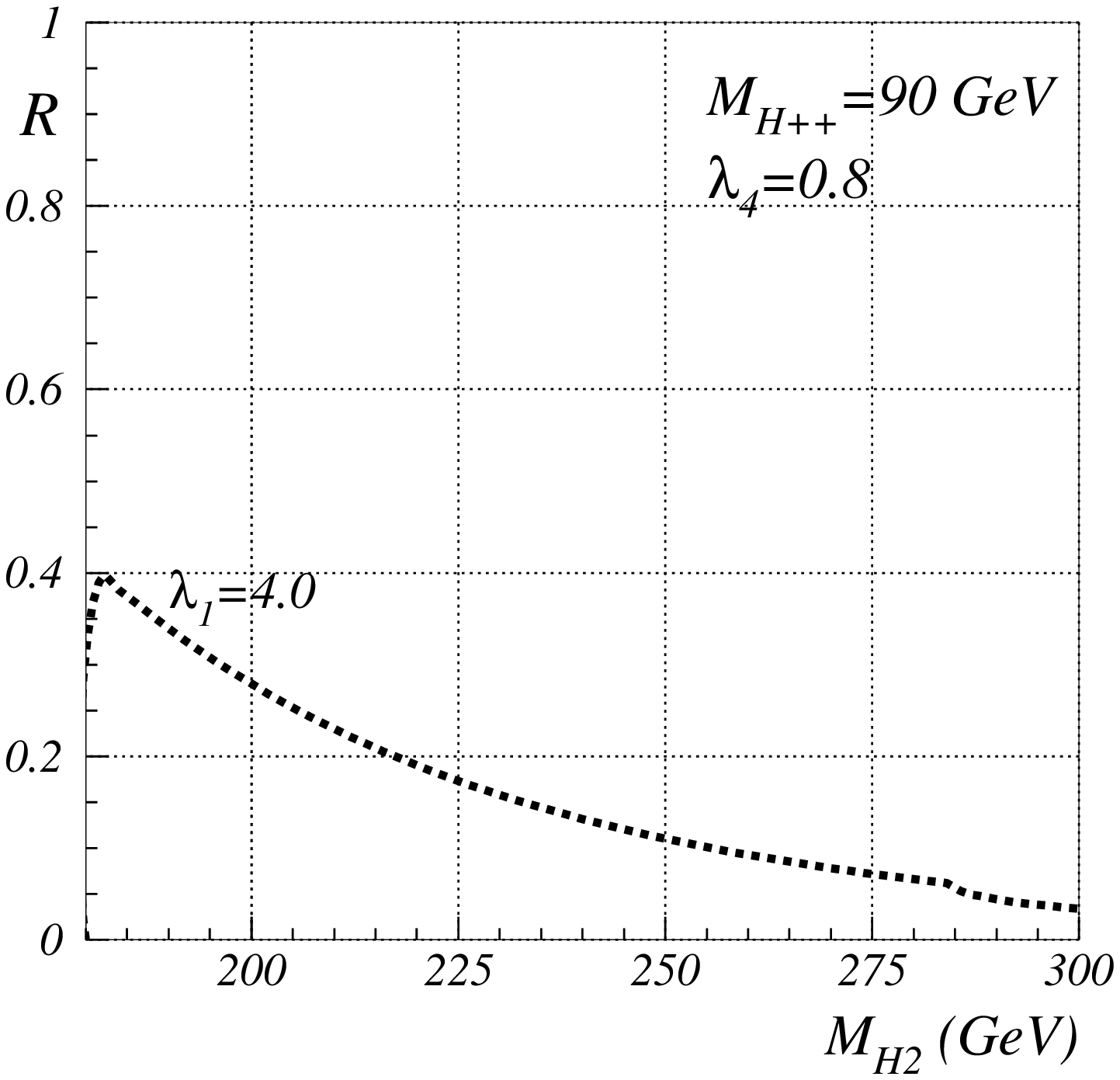}
\includegraphics[origin=c, angle=0, scale=0.46]{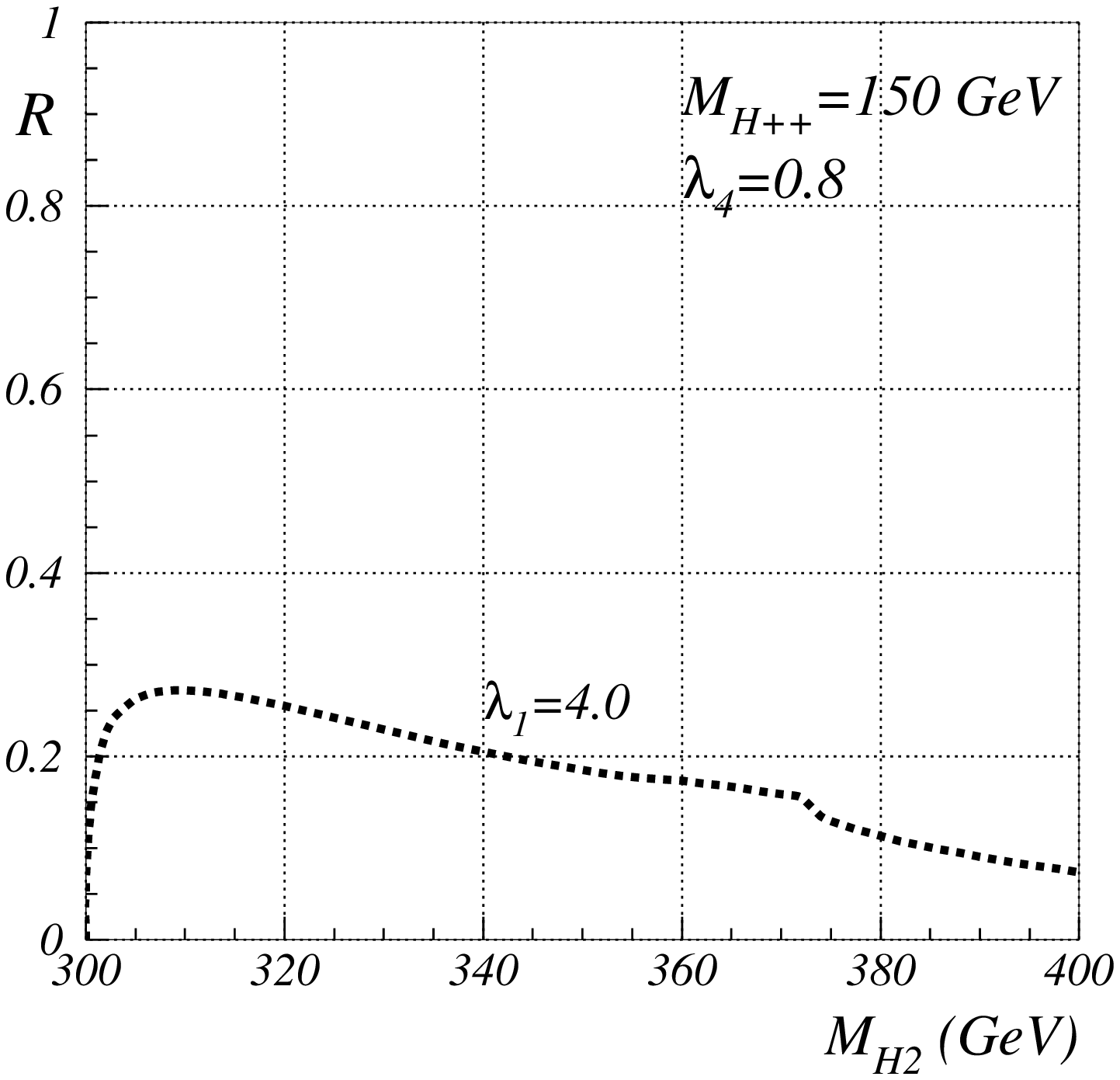}
\caption{The ratio $R=\sigma(gg \to H_2)\times {\rm BR} (H_2\to H^{++}H^{--})/
\sigma(q\overline q\to H^{++}H^{--})$
at the Tevatron (with $\sqrt s=1.96$ TeV) as a function of $M_{H_2}$.
The curves are for $\lambda_1=4$ and $\lambda_4=0.8$ .
We take $M_{H^{\pm\pm}}=90$ GeV in panel (a), $M_{H^{\pm\pm}}=150$ GeV in panel (b).}
\label{fig.7}
\end{center}
\end{figure}

In fig.~(\ref{fig.6}) we plot the analogies of fig.~(\ref{fig.5}) for the LHC with $\sqrt s=7$ TeV. One sees a similar qualitative behaviour,
with lower maximum values of $R$.
In fig.~(\ref{fig.7}) we plot the corresponding results for the Tevatron, for $M_{H^{\pm\pm}}=90$ GeV and $M_{H^{\pm\pm}}=150$ GeV.
Since $\sigma(gg \to H_2)\lsim \sigma(q\overline q\to \gamma^*,Z^*\to H^{++}H^{--})$ for 
$M_{H_2}>2M_{H^{\pm\pm}}$ at the Tevatron, the maximum value of $R\sim 0.4$ (for $\lambda_1=4$) is much smaller than at the LHC and 
is comparable to the QCD K factor for $\sigma(q\overline q\to \gamma^*,Z^*\to H^{++}H^{--})$ \cite{Muhlleitner:2003me}.

Finally, we quantify the number of $H^{++}H^{--}$ events for a given integrated luminosity ${\cal L}$ 
at the LHC. We introduce the parameter $N_{H^{\pm\pm}}$, which is defined as follows:  
\begin{equation}
N_{H^{\pm\pm}}=\epsilon\times {\cal L}\times[\sigma(q\overline q\to \gamma^*,Z^*\to H^{++}H^{--})+
\sigma(gg \to H_2)\times {\rm BR} (H_2\to H^{++}H^{--})]
\label{NH}
\end{equation}
The efficiency $\epsilon$ is the fraction of $H^{++}H^{--}$ events which remain after all 
acceptance/selection cuts are imposed to reduce the background from the SM. The value of $\epsilon$
depends on which decay channel $H^{\pm\pm}\to \ell^\pm_i\ell^\pm_j$ is being considered.
From the LHC simulation in \cite{Rommerskirchen:2007jv} 
for the decay $H^{\pm\pm}\to \mu^\pm\mu^\pm$ with $\sqrt s=14$ TeV, one can derive an approximate value of 
$\epsilon_{\mu\mu}=0.73$ for $M_{H^{\pm\pm}}=600$ GeV and
$\epsilon_{\mu\mu}=0.64$ for $M_{H^{\pm\pm}}=300$ GeV. As expected, 
the efficiency is greater for larger $M_{H^{\pm\pm}}$, since the
leptons originating from $H^{\pm\pm}$ are more energetic.
Extrapolating these values to the region of $M_{H^{\pm\pm}}<300$ GeV (the mass region on which we will focus) one 
roughly obtains
$0.5 < \epsilon_{\mu\mu}<0.6$. The efficiencies for the decay channels $H^{\pm\pm}\to e^\pm e^\pm$ and
$H^{\pm\pm}\to e^\pm\mu^\pm$ are expected to be similar in magnitude to $\epsilon_{\mu\mu}$ 
(see \cite{CMS-search}).
The efficiencies for the decays of $H^{\pm\pm}$ to final states involving 
a $\tau$ lepton are much lower e.g. in \cite{CMS-search} one 
can derive $\epsilon_{\mu\tau}\sim 0.02$ for the channel $H^{\pm\pm}\to \mu^\pm\tau^\pm$,
with even lower values for the channel $H^{\pm\pm}\to \tau^\pm\tau^\pm$.
We will show results for the decay mode $H^{\pm\pm}\to \mu^\pm\mu^\pm$, for 
$M_{H^{\pm\pm}}=90$ GeV, 150 GeV, 200 GeV and 300 GeV.
The strongest lower bounds on $M_{H^{\pm\pm}}$ in this channel 
(assuming a branching ratio of $100\%$) are $M_{H^{\pm\pm}}>156$ GeV from the LHC in
\cite{CMS-search} and $M_{H^{\pm\pm}}> 245$ GeV from the Tevatron in \cite{Aaltonen:2011rt}, both limits being
preliminary results. For the case of BR($H^{\pm\pm}\to \mu^\pm\mu^\pm)< 100\%$,
one can derive from \cite{Aaltonen:2011rt} the approximate limits 
$M_{H^{\pm\pm}}\gsim 100$ GeV, $\gsim 150$ GeV and $\gsim 200$ GeV for BR$\gsim 3\%$,$\gsim 15\%$ and $\gsim 40\%$, respectively. 
We do not include these values of BR($H^{\pm\pm}\to \mu^\pm\mu^\pm$) when showing results for $N_{H^{\pm\pm}}$.
In future searches which require three or four leptons (as done in the LHC search in \cite{CMS-search}) 
the event number $N_{H^{\pm\pm}}$
in eq.~(\ref{NH}) needs to be scaled by a 
multiplicative factor of $[{\rm BR}(H^{\pm\pm}\to \ell^\pm_i\ell^\pm_j)]^2$. As explained above, 
this factor is necessarily less than unity if one considers $M_{H^{\pm\pm}}< 245$ GeV with decay $H^{\pm\pm}\to \mu^\pm\mu^\pm$.

\begin{figure}[t]
\begin{center}
\includegraphics[origin=c, angle=0, scale=0.46]{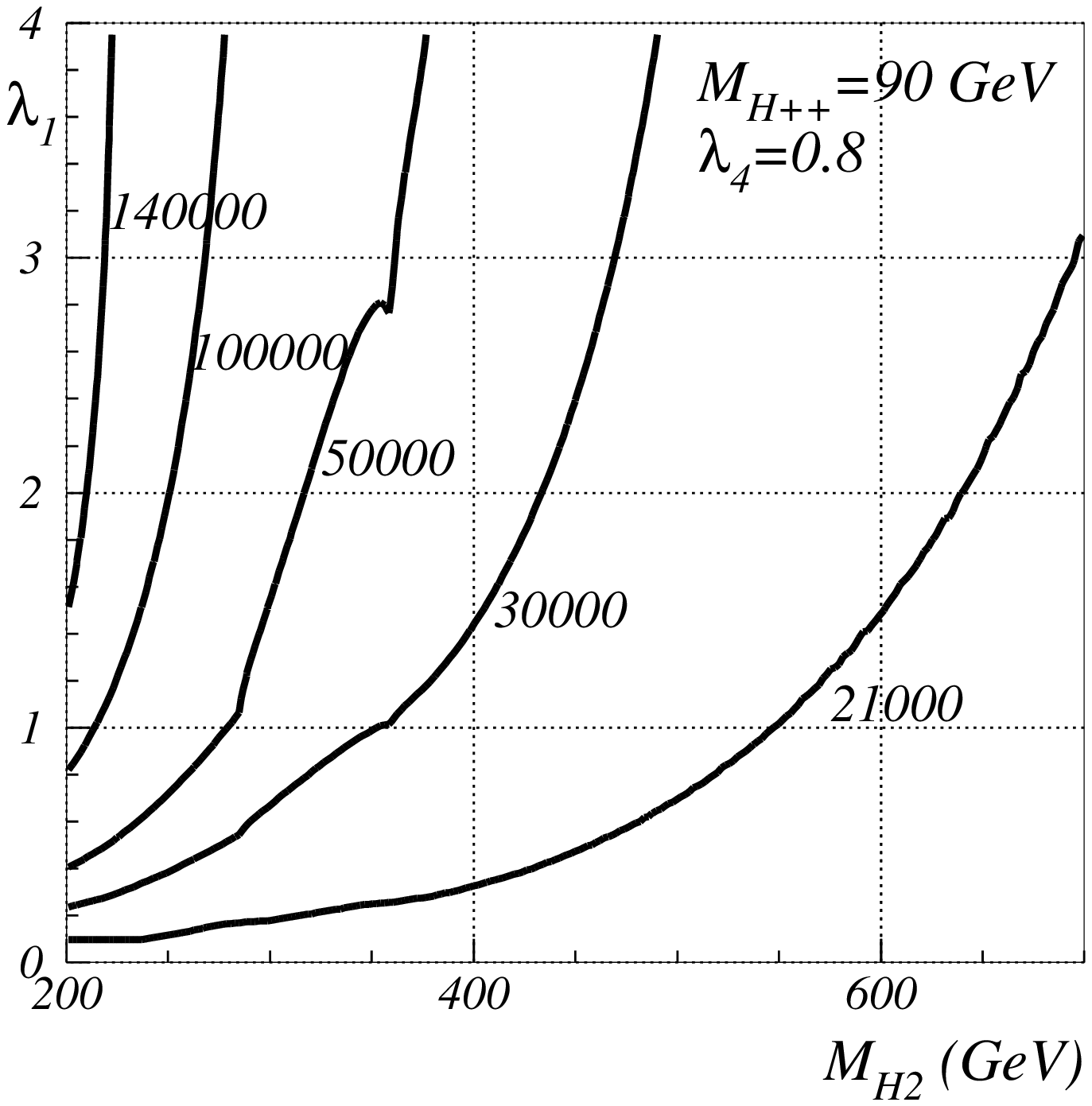}
\includegraphics[origin=c, angle=0, scale=0.46]{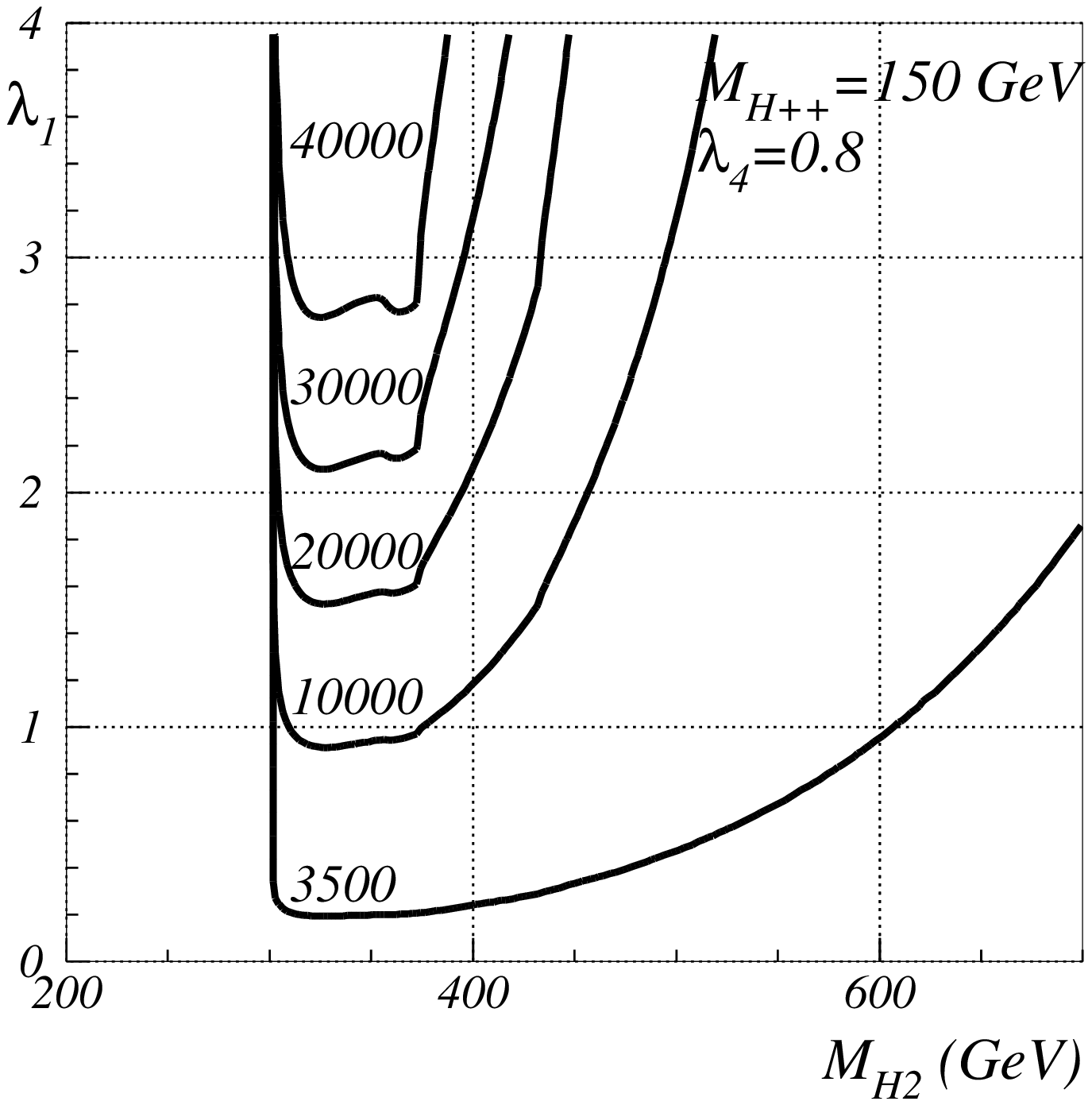}
\includegraphics[origin=c, angle=0, scale=0.46]{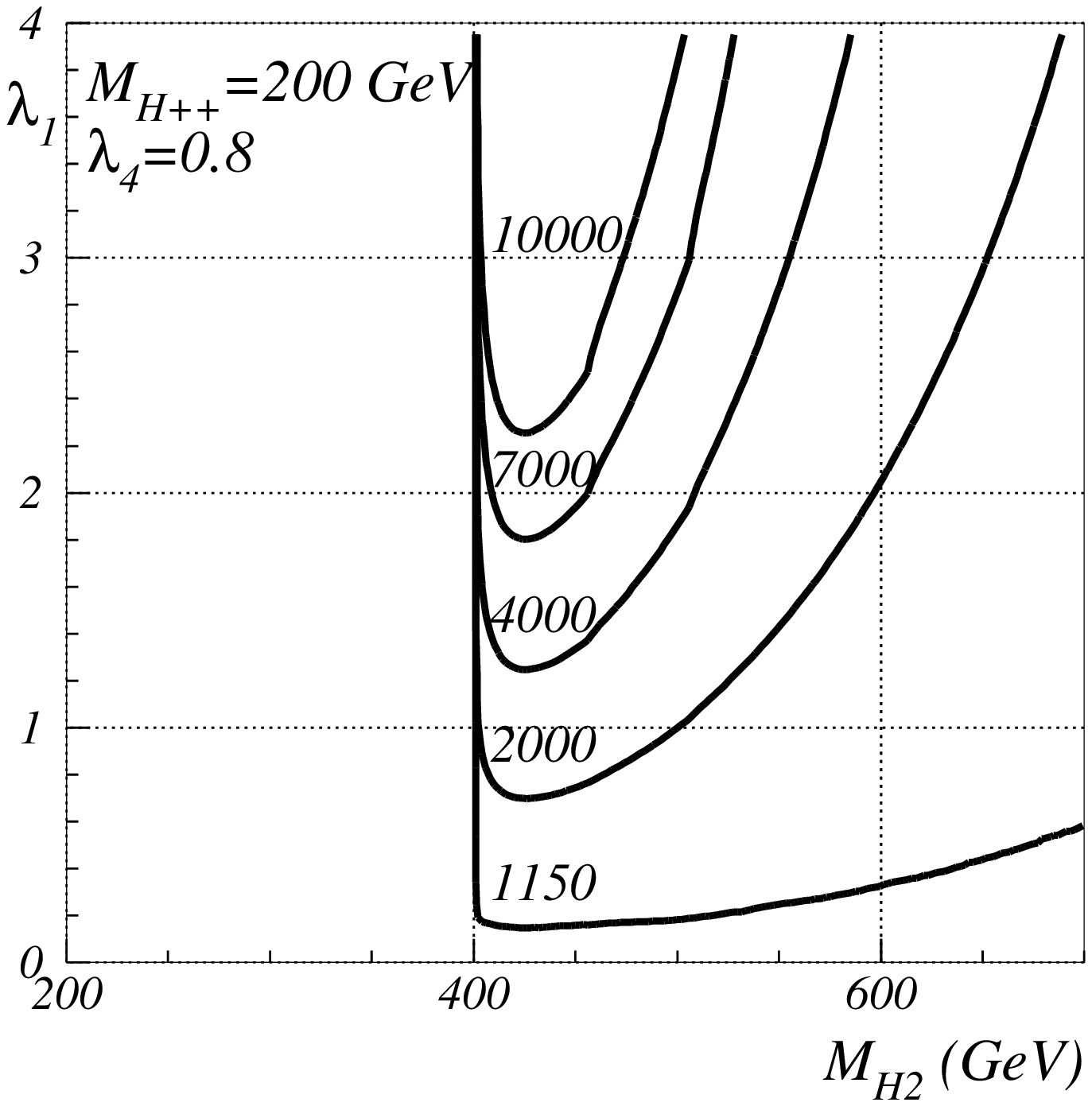}
\includegraphics[origin=c, angle=0, scale=0.46]{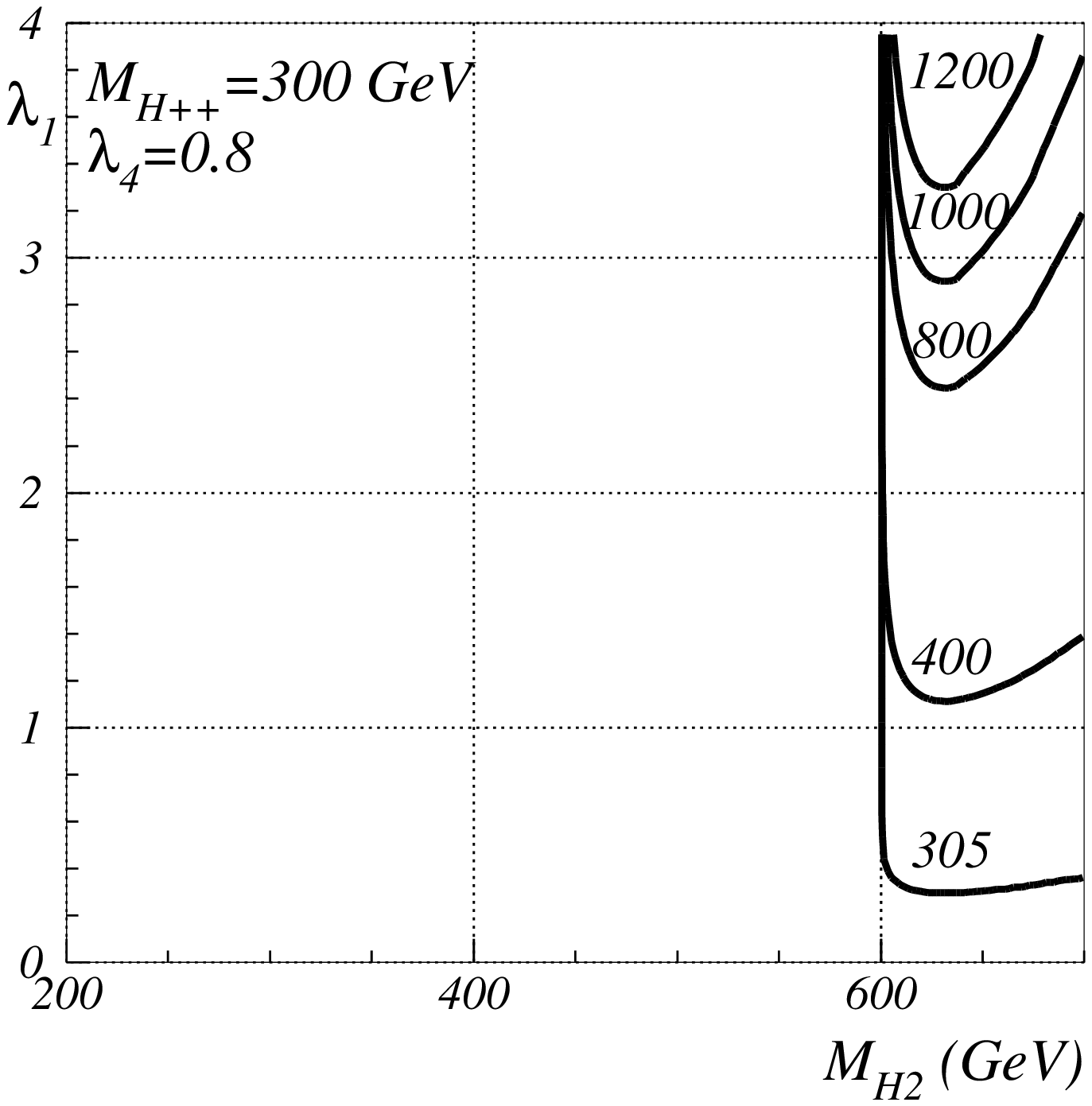}
\caption{The number of $H^{++}H^{--}$ events (assuming BR($H^{\pm\pm}\to \mu^\pm\mu^\pm)=100\%$)
at the LHC with $\sqrt s=14$ TeV and ${\cal L}=30$ fb$^{-1}$ in the plane [$M_{H_2},\lambda_1$].
We take $M_{H^{\pm\pm}}=90$ GeV in panel (a), $M_{H^{\pm\pm}}=150$ GeV in panel (b),
$M_{H^{\pm\pm}}=200$ GeV in panel (c) and  $M_{H^{\pm\pm}}=300$ GeV in panel (d).
In all figures $\lambda_4=0.8$. The number of events for
$q\overline q \to \gamma,Z\to H^{++}H^{--}$ alone
is 20500, 3270, 1130 and 299 
in (a),(b),(c) and (d), respectively.}
\label{fig.8}
\end{center}
\end{figure}

In fig.~(\ref{fig.8}) we show $N_{H^{\pm\pm}}$ in the plane $[M_{H_2},\lambda_1]$
for $\sqrt s=14$ TeV with ${\cal L}$=30 fb$^{-1}$. 
We use $\epsilon_{\mu\mu}=0.64$ for $M_{H^{\pm\pm}}=300$ GeV, and $\epsilon_{\mu\mu}=0.50$ for the other chosen values of
$M_{H^{\pm\pm}}$ (90 GeV, 150 GeV, 200 GeV).
The contribution to $N_{H^{\pm\pm}}$ from $q\overline q \to \gamma^*,Z^*\to H^{++}H^{--}$ alone
does not depend on $M_{H_2}$ and $\lambda_1$, and is roughly equal to 20500, 3270, 1130 and 280
for $M_{H^{\pm\pm}}=90$ GeV, 150 GeV, 200 GeV and 300 GeV respectively. 
In each panel in fig.~(\ref{fig.8}) the contour with the lowest number of events 
corresponds to a value of $N_{H^{\pm\pm}}$ which is slightly larger than  
the above values for $N_{H^{\pm\pm}}$ from $q\overline q \to \gamma^*,Z^*\to H^{++}H^{--}$ alone.
We emphasise that the displayed $N_{H^{\pm\pm}}$ for $M_{H^{\pm\pm}}=90$ GeV, $150$ GeV and $200$ GeV
need to be multiplied by the square of ${\rm BR}$ (for a future three or four lepton search) 
where ${\rm BR}\lsim 3\%$,$\lsim 15\%$ and $\lsim 40\%$
in order to comply with the mass limits in \cite{Aaltonen:2011rt}.
Fig.~(\ref{fig.8})  can also be applied to other decay channels such as $H^{\pm\pm}\to \mu^\pm\tau^\pm$ after multiplying
$N_{H^{\pm\pm}}$ by $\epsilon_{\mu\tau}/\epsilon_{\mu\mu}\sim 1/30$.
Clearly, the contribution from 
$\sigma(gg \to H_2)\times {\rm BR} (H_2\to H^{++}H^{--})$ could significantly enhance the
number of $H^{++}H^{--}$ events at the LHC, provided that $M_{H_2}> 2M_{H^{\pm\pm}}$ and
$\lambda_1$ is not very small. Since it is not expected that $M_{H_2}\gsim 700$ GeV
(from considering constraints from perturbativity and unitarity e.g. see \cite{Arhrib:2011uy}), the
enhancement from $\sigma(gg \to H_2)\times {\rm BR} (H_2\to H^{++}H^{--})$ is limited to 
the region $M_{H^{\pm\pm}}\lsim 350$ GeV. However, its contribution would allow 
the possibility of probing smaller values of ${\rm BR}(H^{\pm\pm}\to \ell^\pm_i\ell^\pm_j)$
for a given value of $M_{H^{\pm\pm}}$ (provided that $M_{H_2}> 2M_{H^{\pm\pm}}$).

\begin{figure}[t]
\begin{center}
\includegraphics[origin=c, angle=0, scale=0.46]{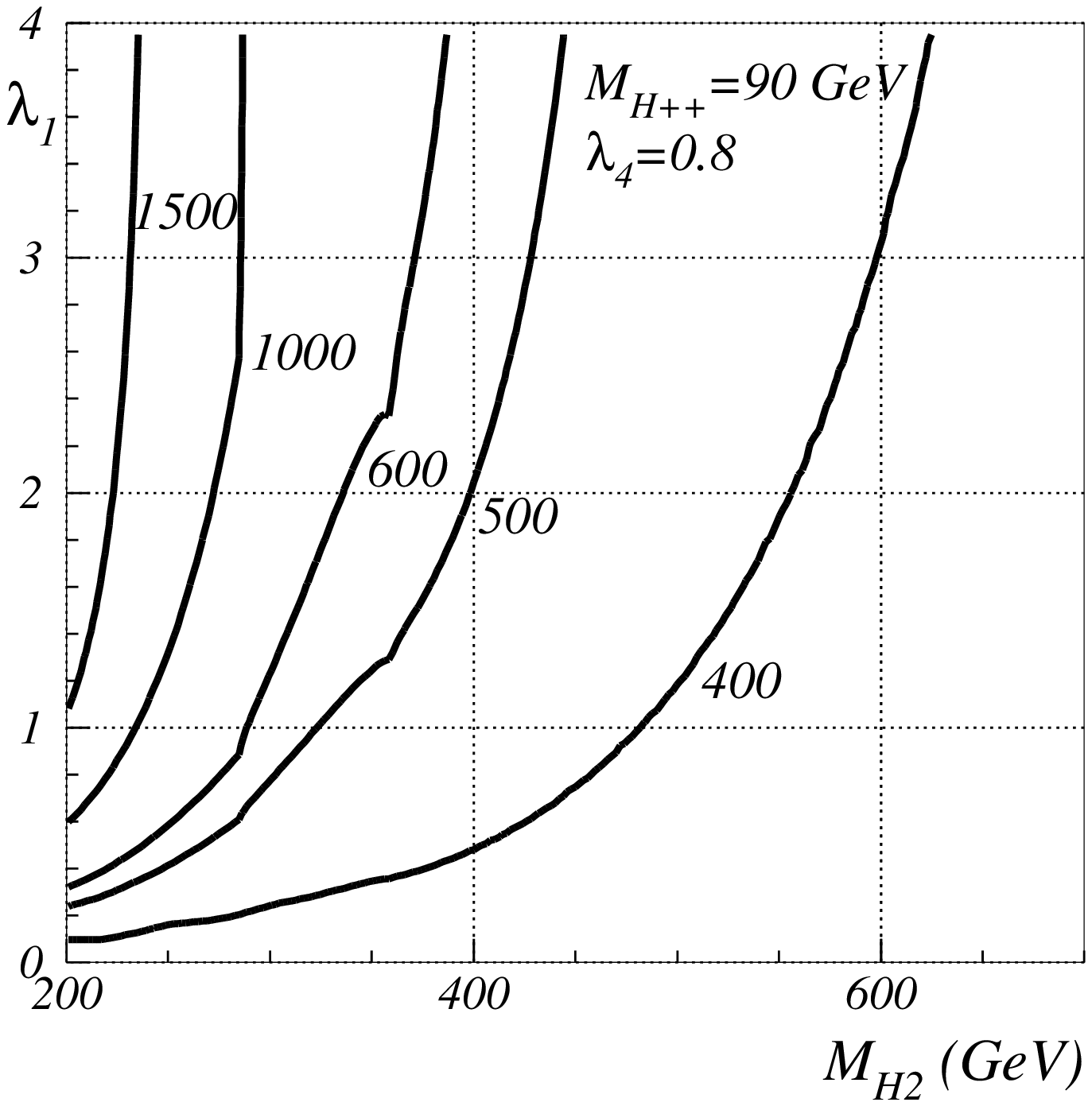}
\includegraphics[origin=c, angle=0, scale=0.46]{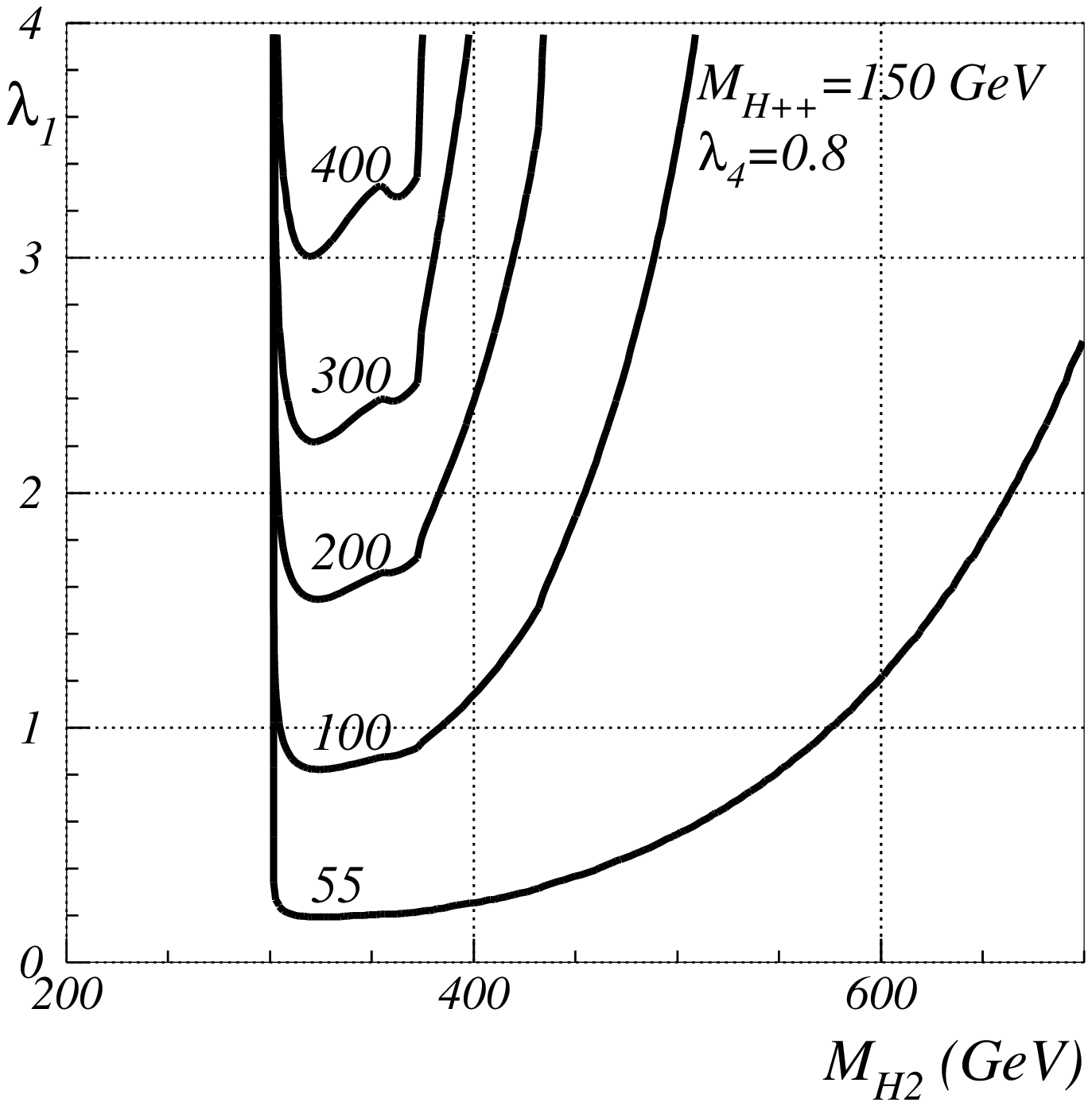}
\includegraphics[origin=c, angle=0, scale=0.46]{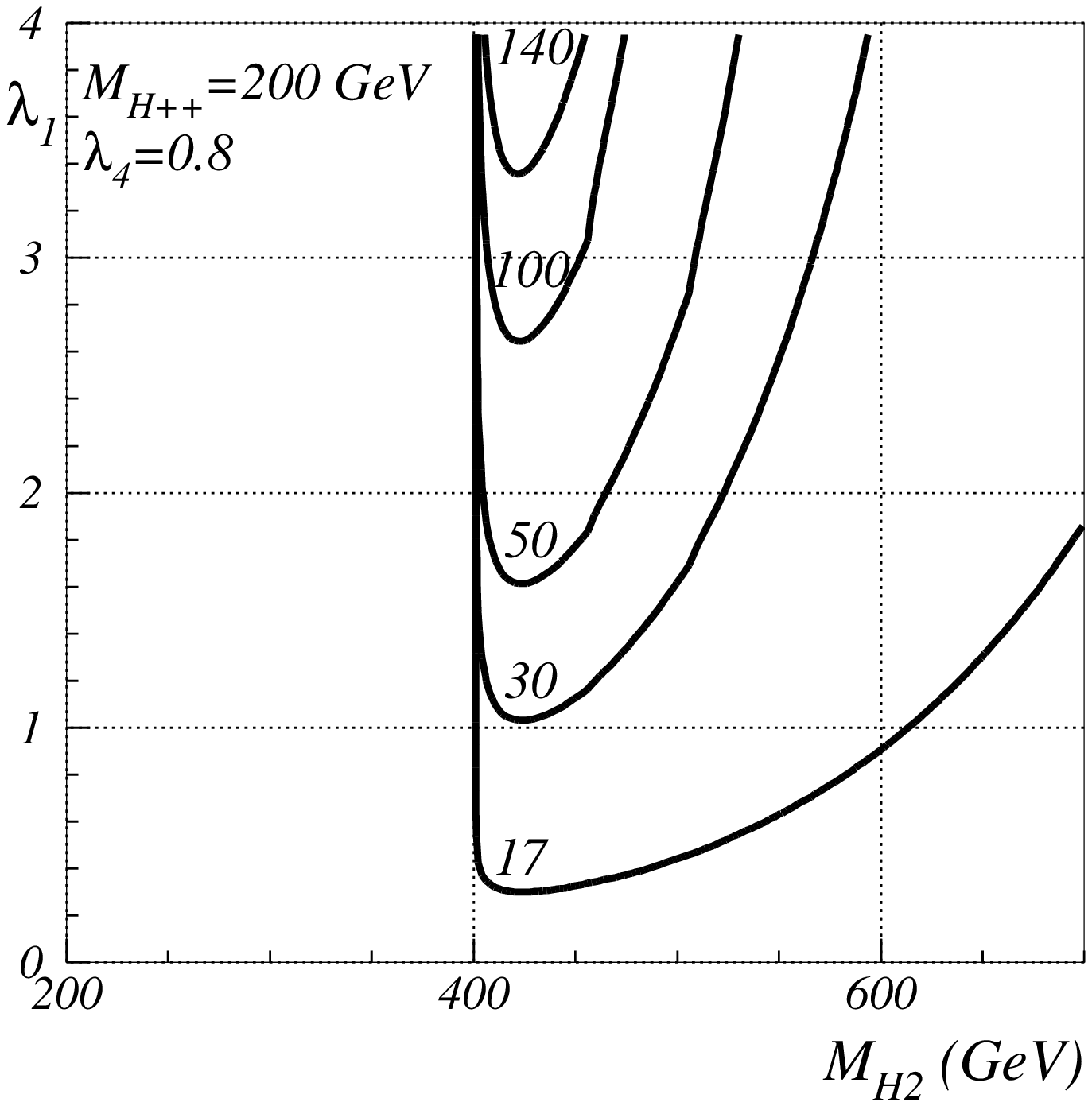}
\includegraphics[origin=c, angle=0, scale=0.46]{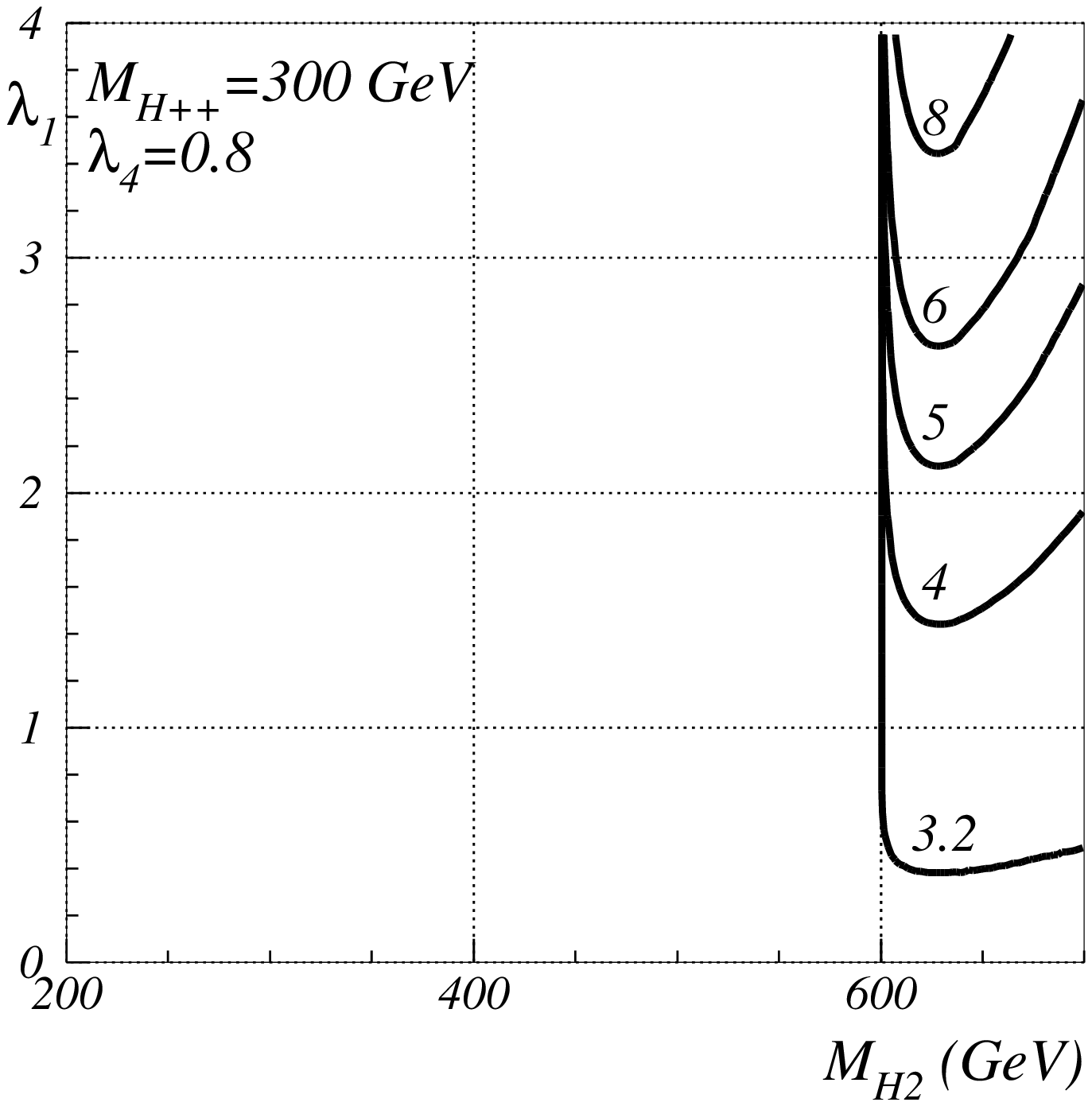}
\caption{The number of $H^{++}H^{--}$ events (assuming BR($H^{\pm\pm}\to \mu^\pm\mu^\pm)=100\%$)
at the LHC with $\sqrt s=7$ TeV and ${\cal L}=2$ fb$^{-1}$ in the plane [$M_{H_2},\lambda_1$].
We take $M_{H^{\pm\pm}}=90$ GeV in panel (a), $M_{H^{\pm\pm}}=150$ GeV in panel (b),
$M_{H^{\pm\pm}}=200$ GeV in panel (c) and  $M_{H^{\pm\pm}}=300$ GeV in panel (d).
In all figures $\lambda_4=0.8$. The number of events for 
$q\overline q \to \gamma,Z\to H^{++}H^{--}$ alone is 390, 53, 16, 3 in 
(a),(b),(c) and (d), respectively.}
\label{fig.9}
\end{center}
\end{figure}

In fig.~(\ref{fig.9}) we show $N_{H^{\pm\pm}}$ for $\sqrt s=7$ TeV with ${\cal L}$=2 fb$^{-1}$. We take
a slightly lower efficiency of $\epsilon_{\mu\mu}=0.4$ for $M_{H^{\pm\pm}}=90$ GeV, 150 GeV and 200 GeV, which is in rough agreement
with the value for the channel $H^{\pm\pm}\to \mu^\pm\mu^\pm$ in the experimental search 
at $\sqrt s=7$ TeV in \cite{CMS-search}. For $M_{H^{\pm\pm}}=300$ GeV we take $\epsilon_{\mu\mu}=0.5$. 
Again, the enhancement from $\sigma(gg \to H_2)\times {\rm BR} (H_2\to H^{++}H^{--})$ can be sizeable,
and could lead to a quicker discovery of a light $H^{\pm\pm}$ with ${\rm BR}(H^{\pm\pm}\to \ell^\pm_i\ell^\pm_j)<100\%$. 
We do not plot an analogous figure for the Tevatron since the maximum value of $R$ is much smaller than at the
LHC, as shown in fig.~(\ref{fig.6}).

Finally, we emphasise that the parameter space of $M_{H_2}> 2M_{H^{\pm\pm}}$
will be probed by two distinct searches with the LHC data taken at $\sqrt s=7$ TeV:
i) the search for
$H_2\to WW,ZZ$ (with first results in \cite{Collaboration:2011qi,Chatrchyan:2011tz}), and ii) the search for 
$q\overline q \to \gamma^*,Z^*\to H^{++}H^{--}$ (with first results in \cite{CMS-search}, and a simulation
in \cite{Rentala:2011mr}). The run at $\sqrt s=7$ TeV, with possibly up to 5 fb$^{-1}$ of integrated luminosity,
has the potential to exclude or provide evidence for a SM-like Higgs boson with $200 \,{\rm GeV} < M_{H_2}< 500$ GeV at a high confidence
level. Therefore the scenario of $M_{H_2}> 2M_{H^{\pm\pm}}$ and its possible impact on the
direct searches for $H^{\pm\pm}$ should be clarified within the next two years.

\section{Conclusions}
Doubly charged Higgs bosons ($H^{\pm\pm}$), which arise in the Higgs Triplet Model (HTM)
of neutrino mass generation, are being searched for at the Tevatron and at the LHC.
The ongoing searches assume the production mechanisms $q\overline q \to \gamma^*,Z^*\to H^{++}H^{--}$
and $q'\overline q \to W^* \to H^{\pm\pm}H^{\mp}$, with the leptonic decay  $H^{\pm\pm}\to \ell^\pm_i\ell^\pm_j$.
We proposed an additional production mechanism for $H^{\pm\pm}$, which becomes possible
if the SM-like Higgs boson in the HTM ($H_2$) is heavy enough to decay to a pair of on-shell $H^{\pm\pm}$. 
We quantified the magnitude of the branching ratio of $H_2\to H^{++}H^{--}$, and showed that
it can be large ($>> 10\%$) if a quartic coupling in the scalar potential is sizeable, $\lambda_1> 1$.
We performed a numerical study of the production rate for $H_2$ via gluon-gluon fusion, 
$gg\to H_2$, followed by the decay $H_2\to H^{++}H^{--}$,
and we showed that its cross section at the LHC can be greater than 
that of $q\overline q \to \gamma^*,Z^*\to H^{++}H^{--}$ in
a sizeable parameter space of $[\lambda_1,M_{H_2}]$ (with $M_{H_2}> 2M_{H^{\pm\pm}}$).
In the optimal case (e.g. $\lambda_1=4$, $M_{H_2}\sim 420$ GeV and $M_{H^{\pm\pm}}=200$ GeV)
the ratio of cross sections can be as large as nineteen. 
We note that our analysis was carried out using the leading-order cross sections
only, and the inclusion of QCD $K$ factors would provide a further enhancement of 40\% in the above ratio. Such an additional source of 
$H^{++}H^{--}$ would enable smaller values of the branching ratio of $H^{\pm\pm}\to \ell^\pm_i\ell^\pm_j$ to be probed
at the LHC. The case of $M_{H_2}> 2M_{H^{\pm\pm}}$ necessarily requires $M_{H_2}\gsim 200$ GeV,
and this mass region is now being probed for the first time at the LHC for the decay channels of $H_2$ to SM particles, 
$H_2\to WW$ and $H_2\to ZZ$. The possibility of a SM-like Higgs boson in the HTM with $M_{H_2}> 2M_{H^{\pm\pm}}$ 
and its potential impact on the direct searches for $H^{\pm\pm}$ should be clarified within the $\sqrt s=7$ TeV run at the LHC.

\section*{Acknowledgements}
We thank Hiroaki Sugiyama and Abdesslam Arhrib for useful discussions. A.G.A was supported by a Marie Curie 
Incoming International Fellowship, FP7-PEOPLE-2009-IIF, Contract No. 252263.  
This work is supported in part by the NExT Institute.

\section*{Note Added}
After submission of this paper, the LHC searches for the SM Higgs boson were
updated with ${\cal L} =1.1$ fb$^{-1}$ \cite{eps:Murray}. For the region $M_{H_2}> 200$ GeV,
both of the CMS and ATLAS collaborations use the decay channel $H_2\to ZZ$, with subsequent decays
$ZZ\to \ell^+\ell^-\nu\nu, ZZ\to \ell^+\ell^-q\overline q$ and $ZZ\to \ell^+\ell^-\ell^+\ell^-$.
CMS also search for $H_2\to WW$ with the decay mode $WW\to \ell\nu\ell\nu$, while ATLAS
search for $H_2\to WW$ with the decay mode $WW\to \ell\nu q'\overline q$.
After combining the results from these four distinct channels, both collaborations 
exclude at 95\% c.l the mass range $295 \,{\rm GeV} < M_{H_2}< 450 \,{\rm GeV}$.
This does not preclude a sizeable value of $R$ in the HTM,  e.g. from fig.~\ref{fig.6}c, one can see 
that $7 > R > 1$ in the interval $450 \,{\rm GeV} < M_{H_2}< 600 \,{\rm GeV}$, for $\lambda_1=4$ and
$\sqrt s=7$ TeV (and not including the enhancement from the QCD $K$ factor).

\end{document}